\documentclass[conference]{IEEEtran}
\usepackage{graphicx}
\usepackage{amsmath,amssymb}
\usepackage[numbers,sort&compress]{natbib}
\usepackage{url}
\usepackage[breaklinks=true,hidelinks]{hyperref}
\makeatletter
\g@addto@macro{\UrlBreaks}{\UrlOrds}
\makeatother
\usepackage{lmodern}
\usepackage{listings}
\usepackage{xcolor}
\usepackage{tabularx}
\usepackage{booktabs}
\usepackage[table]{xcolor}
\usepackage{booktabs}
\usepackage{float}
\definecolor{heatblue}{RGB}{49,130,189}

\graphicspath{{./}{./figures/}}

\usepackage{eso-pic}
\newcommand{\PreprintNotice}{%
  \AddToShipoutPictureFG*{%
    \AtPageLowerLeft{%
      \hspace*{5mm}\raisebox{5mm}{%
        \parbox{0.95\paperwidth}{%
          \footnotesize\sffamily\color{black}%
          Accepted at the 33rd IEEE Symposium on High-Performance Interconnects (HotI 2026). Lightmatter.
          This work has been submitted to the IEEE for possible publication.
          Copyright may be transferred without notice, after which this version may no longer be accessible.}}}}}

\author{Arulselvan Madhavan, Peter Carson, Taylor Groves, Thomas Graham}
\date{\today}
\title{Scaling Inference Prefill with High-Radix Photonic Interconnects}
\hypersetup{
 pdfauthor={Arulselvan Madhavan, Peter Carson, Taylor Groves, Thomas Graham},
 pdftitle={Scaling Inference Prefill with High-Radix Photonic Interconnects},
 pdfkeywords={Silicon Photonics, AI Inference, Photonic Interconnects, Large Language Models},
 pdfsubject={cs.DC},
 pdflang={English}}

\begin{document}

\PreprintNotice
\maketitle

\begin{abstract}
With the rise of inference as today's dominant AI workload, the industry is transitioning to high-bandwidth photonic interconnects to meet the large scale-up requirements of increasingly complex Mixture-of-Experts (MoE) models. This paper quantifies the benefits of 3D-integrated photonic interconnects for inference prefill by analyzing tradeoffs between high-concurrency throughput for Large Language Model (LLM) chat and the large context windows typically required for reasoning and agentic AI. We simulate three MoE models: short context (1K--8K tokens), medium context (128K tokens), and long context (1M tokens). We evaluate this workload across existing copper-based GPU systems and one with high bandwidth integrated photonics. We show 2.1--3.2$\times$ latency improvements in the stressed high-batch regimes and 2.8--5.8$\times$ improvements over baselines in communication-limited configurations. 3D photonics enable the 1152-GPU footprint required to lower time-to-first-token, yielding 2.2--4.5$\times$ speedups across production-grade platforms when electrical systems cross their inherent scale-up-pod limits.
\end{abstract}


\section{Introduction}
\label{sec:introduction}

The proliferation of AI-driven applications and broad global adoption have transformed the compute infrastructure landscape; while frontier training remains critical, the industry has matured into an inference-first era, with inference now accounting for 80--90\% of compute cycles \citep{Tensormesh2025,Fullview2025}. The rise of reasoning models and agentic workflows demands "continuous inference" capabilities, forcing providers to balance the high-concurrency throughput required for LLM chat services against the large context windows required for deep research and agentic AI. In production agentic coding workflows, the median initial context (prefill) length already reaches ${\sim}$96K tokens---driven by accumulated tool-call outputs, system prompts, and skill primers---with nearly half of all requests exceeding 128K tokens \citep{semianalysis2026tokens}, making prefill the dominant computational phase. This not only places simultaneous pressure on system latency and memory bandwidth but also creates a fundamental engineering dilemma in system optimization between cost-efficiency and user experience.

The trajectory of AI hardware follows the path of domain-specific architecture (DSA) evolution described by Patterson and Hennessy \citep{hennessy2019golden}, moving toward silicon increasingly specialized for the distinct phases of LLM execution. This specialization has led to a bifurcation in silicon design: for example, NVIDIA’s Rubin roadmap focuses on large parallel compute particularly suited for the prefill phase of inference, while architectures like the Groq Language Processing Unit (LPU) prioritize high-speed, deterministic memory access to optimize the decode phase.

However, these refinements do not eliminate system bottlenecks; they disperse and transform them. While raw floating-point operations per second (FLOPS) continue to scale, the "choke point" for decode remains the memory bandwidth \citep{pope2023efficiently}. For inference prefill, the reach and shoreline limits of conventional interconnects are the primary constraint for large-scale distributed systems. In the prefill phase, providers face a fundamental batching dilemma: increasing batch sizes is necessary to maximize GPU utilization and aggregate throughput. However, larger batches proportionately increase the volume of data exchanged in distributed collectives---tensor-parallel all-reduces, expert-parallel all-to-alls, and sequence-parallel all-gathers---across the compute cluster \citep{korthikanti2022reducing}. This creates a major networking challenge in the form of large activation tensors continuously exchanging large amounts of data across the compute cluster.

With electrical interconnects, which are constrained by a fundamental bandwidth-reach-power envelope, the communication overhead of these workloads becomes a primary bottleneck. As per-lane signaling rates increase to satisfy bandwidth demands, the effective reach of passive copper shrinks to 1m at 224G and to tens of centimeters at 448G \citep{lm2025moetraining}, effectively restricting high-performance scale-up pods to a single rack, forcing the use of slower scale-out fabrics for multi-rack communication. 

In distributed inference, communication overhead can account for more than half of total prefill latency---over 65\% on bandwidth-limited inference GPUs under tensor parallelism \citep{li2024flashcommunication}---making interconnect bandwidth a first-order constraint on inference throughput and latency. Conversely, in reasoning and agentic scenarios (128K to 1M+ tokens), meeting strict time-to-first-token (TTFT) service level agreements (SLAs) requires extreme parallelism across multiple racks encompassing hundreds of nodes. Photonic interconnects enable the expansion of the high-bandwidth scale-up pod to hundreds or thousands of compute packages across multiple racks. This allows inference engines to sustain larger batch sizes for throughput and large context for reasoning while maintaining compute utilization.

In this work, we quantify the impact of 3D-integrated photonics on AI inference efficiency. We demonstrate how 4x bandwidth and radix up to 1,152 unlock new opportunities for larger context handling at higher batch sizes, specifically detailing:

\begin{itemize}
    \item \textbf{High Bandwidth Impact:} Increased bandwidth between compute nodes allows inference engines to maintain large-scale batch sizes while meeting SLA latency constraints.
    \item \textbf{Expanded Scale-Up Pods:} Extending high-bandwidth scale-up pods up to 1152 devices ensures the ability to process reasoning contexts (128K+ tokens) without incurring scale-out penalties.
    \item \textbf{Large Context Viability:} For emerging 1M+ token windows, the bandwidth and scale-up pod size achievable with optical interconnect technology are necessary to achieve fast TTFT metrics.
\end{itemize}

Overall, on the NVIDIA B300 platform with FP4 quantization and using a 42B-active-parameter MoE model, optical interconnects deliver approximately 2.6$\times$ improvement in prefill latency for 1K-token contexts, 2.2$\times$ for 8K, 2.9$\times$ for 128K, and 2.3$\times$ for 1M tokens when looking at systems comprising 72 to 1152 processors (Table~\ref{tab:sweep_heatmap}). These ratios depend on the operating regime---defined by batch size, context length, and device count relative to the native scale-up pod---and should not be read as universal improvements: configurations that are compute-bound show modest gains, whereas those that push batch size, accelerate per-GPU compute, or cross the scale-up pod boundary expose larger communication penalties and benefit more from higher interconnect bandwidth. Rack-limited B200 and Rubin baselines show larger improvements at 128K--1M input tokens because communication and scale-up pod limits dominate prefill performance. Baselines with larger hybrid electrical-and-optical scale-up pods reduce the comparative improvement at 128K input tokens, but the all-optical configuration still unlocks substantially larger gains for 1M tokens.

\section{Background}
\label{sec:background}

\subsection{LLM Inference Workload Characteristics}
\label{subsec:inference_workload}
LLM inference is characterized by a distinct dual-phase execution pattern, each with divergent resource demands.

\subsubsection{Prefill Phase (Compute-Bound)}
The prefill phase processes the entire input prompt in parallel to generate the initial Key-Value (KV) cache. Because the full sequence is known a priori, this phase relies on dense matrix multiplications that can saturate GPU compute utilization \citep{agrawal2024sarathi}. The primary performance metric is TTFT, which is directly sensitive to the raw FLOPS the processor can achieve.

\subsubsection{Decode Phase (Memory-Bound)}
Following prefill, the system enters the decode phase to auto-regressively generate output tokens one at a time. Each step requires loading the entire model weights and the active KV cache from memory to produce a single token. This operation has low arithmetic intensity and is fundamentally bound by High Bandwidth Memory (HBM) bandwidth rather than compute capacity \citep{pope2023efficiently}. The critical metric is Time Per Output Token (TPOT), which determines the perceived fluidity of generation.

\subsection{The Efficiency vs.\ Interactivity Challenge}
\label{subsec:efficiency_interactivity}
To achieve high-performance LLM serving, system engineers must navigate a complex landscape of hardware constraints and user expectations.

\subsubsection{Batch Throughput vs. Response Time}
The fundamental tension in inference serving lies in batching. Increasing batch sizes maximizes aggregate system throughput by saturating GPU compute resources, which improves cost efficiency for the provider. However, larger batches simultaneously degrade individual inter-token latency for the user due to increased queuing and iteration times \citep{agrawal2024sarathi}. Providers are thus forced to compromise on throughput to meet strict interactivity SLAs.

\subsubsection{Cache Pressure and Eviction}
Compounding the batching dilemma is memory pressure. Retaining long-context KV caches is essential for seamless user experiences in multi-turn dialogues or reasoning tasks. However, the large memory footprint of these caches limits the total number of concurrent users the hardware can support \citep{pope2023efficiently}. When HBM capacity is exhausted, the system must trigger performance-degrading evictions or offload data to slower CPU memory, creating a ``memory wall'' that caps scalability \citep{continuum2025}.

\subsection{Hardware Bottlenecks in the Scale-Up Pod}
\label{subsec:hardware_limits}
While algorithmic innovations such as continuous batching \citep{agrawal2024sarathi} and disaggregated inference \citep{zhong2024distserve} attempt to mitigate the hardware constraints, system performance is ultimately limited by the physical characteristics of the underlying hardware interconnects \citep{lm2025moetraining}.

\subsubsection{The Shoreline Constraint}
As computational capability scales with die area, I/O bandwidth is constrained to the chip's perimeter, or ``shoreline'' \citep{lm2025moetraining}. Large portions of this shoreline are reserved for HBM to ensure signal integrity, significantly reducing the available area for scale-up interconnect serializer/deserializer (SerDes) links \citep{lm2025moetraining}. This bottleneck restricts the aggregate bandwidth available to support the large batch sizes required for efficiency.

\subsubsection{Reach and Energy Limitations}
Current scale-up fabrics rely on electrical copper interconnects, but the physics of high-speed electrical transmission imposes a ``reach-energy'' trade-off. At 224~Gb/s, the maximum reach of passive electrical interconnects shrinks to approximately 1 meter \citep{lm2025moetraining}. Extending this reach requires power-hungry retimers that divert power away from computation \citep{lm2025moetraining}. Consequently, electrical scale-up pods are effectively limited to a single rack (approximately 72 to 144 GPUs), constraining the size of the logical GPU available for inference \citep{lm2025moetraining}. Scale-out fabrics (e.g., InfiniBand) connect far more devices, but at much lower per-GPU bandwidth (Table~\ref{tab:hw_all}, scale-up (SU) bandwidth vs.\ scale-out (SO) bandwidth), so they do not substitute for a larger scale-up domain.

\subsubsection{ 3D Integrated Photonics}
To overcome these barriers, 3D-integrated photonic interconnects offer a solution by stacking SerDes or host ASICs directly onto the optical engines \citep{lm2025moetraining}. This approach decouples I/O from the package shoreline, allowing SerDes to be distributed across the chip area. This architecture enables a massive increase in bandwidth density and radix, allowing for scale-up pods of greater than 1152 GPUs while mitigating the power and cooling challenges inherent in dense clusters \citep{lm2025moetraining}.

\section{Meeting AI Inference Demands with Semiconductor Design}
\label{sec:semiconductor}

Recently, the market has seen the introduction of solutions tailored towards the specific needs of inference workloads. Major silicon providers such as Nvidia are moving away from general-purpose, one-size-fits-all silicon designs towards architectures targeting disaggregated inference, such as Rubin CPX \citep{nvidia2025rubin} and Groq LPX \citep{groq2024lpu}. Other specialized silicon providers, such as Cerebras and d-Matrix, have built silicon optimized for inference \citep{cerebras2024wse3,dmatrix2024}. These architectures generally aim to balance the ratios of compute power, memory capacity, memory bandwidth, and interconnect bandwidth to match the needs of the prefill or decode phase (or both) of inference.

While compute, memory capacity, and memory bandwidth continue to scale \citep{reuther2023ai}, a pivotal shift in interconnect technology is required to increase bandwidth density, overcome the reach limitations of copper, and build more powerful inference systems.

\subsection{3D Photonic Interconnects}

To address these limitations, the industry is moving toward true 3D integrated photonics, exemplified by the Lightmatter Passage platform \citep{lightmatterpassage}, which we cite only as a reference point for the bandwidth, radix, and energy figures used below (Section~\ref{subsec:hw_config}) and do not directly simulate. Passage utilizes a 3D package architecture where the Electronic Integrated Circuit (EIC) or host ASIC is stacked directly on top of the Photonic Integrated Circuit (PIC).  Passage offers distinct advantages for next-generation inference hardware:
\begin{itemize}
    \item \textbf{Extreme Bandwidth:} By placing optical microring modulators (MRMs) just 100 microns beneath the electrical SerDes, Passage eliminates the need for power-hungry equalization and long electrical traces. This vertical integration allows I/O to be distributed anywhere across the chip area—represented as a ``grid'' of vertical connections—rather than being constrained to the package shoreline. MRMs combine multiple wavelengths (lambdas) of light onto a single fiber. 3D stacking and wavelength division multiplexing enables unprecedented bandwidth. While current GPU configurations, such as the NVIDIA Blackwell architecture, are equipped with 14.4~Tb/s of bidirectional interconnect bandwidth \citep{nvidiaGB200NVL72}, a Passage-enabled GPU can support greater than 64~Tb/s of bidirectional bandwidth per GPU. 
    \item \textbf{High Radix:} With photonic interconnects, data escapes the host GPU or switch using tightly spaced optical fibers. Fibers are 127~{\textmu}m (compared to 254~{\textmu}m for 30AWG copper) and combine both transmit and receive signals on a single fiber.  A bidirectional fiber replaces four copper wires (a differential pair (DP) for transmit and DP for receive).  This allows for 8X increase in radix vs copper-based approaches. Furthermore, optical interconnects overcome the reach limitations of copper cables, allowing for scale-up pods to extend beyond a single rack. This enables both larger scale-up pods and sparser racks, which alleviates power density and cooling constraints. 
    \item \textbf{Energy Efficiency:} For inference workloads where power efficiency is paramount, 3D integrated photonics dramatically reduce the energy cost of data movement. The complete link budget—including the PIC, laser, and host SerDes—operates at approximately 4.3~pJ/bit (2.3 pJ/bit for PIC and laser only), comparable to the $\sim$5~pJ/bit of passive copper and significantly lower than the $>$20~pJ/bit of traditional pluggable optics \citep{lm2025moetraining}.
\end{itemize}

\begin{table}[h]
\centering
\caption{Comparison of Passage optics compared to passive copper.}
\label{tab:passagevcopper}
\resizebox{\columnwidth}{!}{%
\begin{tabular}{l l l c}
\toprule
                       & \textbf{Copper}       & \textbf{3D Optics}                          & \textbf{Improvement} \\
\midrule
Reach (m)              & $\sim$1 (224G)        & 1000                                        & $1000\times$ \\
Port Pitch ({\textmu}m)    & 1020, 30AWG, 2~DP    & 127 bidirectional fiber                    & $8\times$    \\
Tbps/Fiber or Wire     & 0.224/DP              & 1.79, 16$\lambda$                           & $8\times$    \\
Power (pJ/b)           & 5 (SerDes)            & 4.3 (2.3 PIC \& Laser, 2 SerDes)           & $0.13\times$ \\
\bottomrule
\end{tabular}}
\end{table}
Table~\ref{tab:passagevcopper} shows an overview of the benefits of 3D photonics compared to passive copper. In this paper, we quantify the benefits of these photonic interconnects in the AI inference landscape.

\section{Methodology}
\label{sec:methodology}

\subsection{Modeling LLMs}
\label{subsec:modeling_llms}
We model the inference of production-grade Machine Learning (ML) models by generating a Multi-Level Intermediate Representation (MLIR), capturing the model partitions (Tensor parallel, Expert parallel, Context parallel). There are two main reasons for engaging in this style of modeling. First, we want to take implementations deployed in production, with model partitions faithfully captured. Second, our focus is on accurately modeling the communication calls. We have the compiler generate these calls, thereby ensuring their implementation is as it would be in a production system. We maintain a fork of the XLA (Accelerated Linear Algebra) compiler \citep{openxlagithub} that we use to walk through the MLIR file and capture the compute and communication costs of any model.

\subsection{Model Configurations}
\label{subsec:model_config}

We evaluate three model scales representative of MoE frontier models, as shown in Table~\ref{tab:models}. All three models use Multi-Head Latent Attention (MLA) \citep{deepseekv3}, which significantly reduces KV cache size through low-rank compression compared to Grouped-Query Attention (GQA)-based architectures.

\begin{table}[h]
\centering
\caption{Model Configurations Evaluated. R1 denotes a divisibility-adjusted variant of DeepSeek-R1; see Section~\ref{subsec:model_config}.}
\label{tab:models}
\begin{tabularx}{\columnwidth}{lXXX}
\toprule
\textbf{Metric} & \textbf{Mini} & \textbf{R1} & \textbf{Next} \\
\midrule
Layers          & 30   & 61   & 120  \\
Experts         & 144  & 288  & 1152 \\
Attention heads & 144  & 144  & 288  \\
Total params    & 190B & 840B & 13T  \\
Active params   & 21B  & 42B  & 201B \\
\bottomrule
\end{tabularx}
\end{table}

The Mini model (21B active parameters) represents the smallest MoE tier; R1 (42B active parameters) is aligned with the DeepSeek-R1 architecture \citep{deepseekv3}; Next (201B active parameters) represents a larger-scale MoE system. We evaluate both FP4 and FP8 quantization precisions \citep{ieee7542019}. The R1 variant used in this work applies divisibility-motivated parameter adjustments (embedding dimension 8064 compared to 7168 in the production DeepSeek-R1 \citep{jaxllmexamples}, 288 routed experts compared to 256, vocabulary 145 440 compared to 129 280) to satisfy mesh-divisibility constraints across device counts of 8, 16, 24, 48, and 72. This yields a model approximately 14\% larger than production DeepSeek-R1 in parameter count while preserving the same communication topology and scaling behavior. Complete configuration details are in Appendix~\ref{sec:model_impl}.

\subsection{Hardware Configuration}
\label{subsec:hw_config}

We model four GPU architectures at the per-GPU and scale-up-pod specifications summarized in Table~\ref{tab:hw_all}. Scale-up (SU) is the maximum electrical scale-up pod (GPU count); SU bandwidth (BW) is the unidirectional scale-up interconnect bandwidth (GB/s) per GPU across that pod. Scale-out (SO) bandwidth is the unidirectional cross-rack interconnect bandwidth (GB/s) per GPU modeled for devices beyond the SU pod. \textbf{Rubin} denotes a reported roadmap-class Rubin configuration, corresponding to the proposed dual-die \emph{Vera Rubin} GPU with up to 72 GPUs in a scale-up pod (144 logical dies). \textbf{R4} denotes a speculative roadmap-class configuration inferred from the SemiAnalysis report; it models a potential quad-die GPU by extrapolating from that dual-die design, with up to 576 GPUs in a scale-up pod (2304 logical dies) \citep{semianalysis2025inferencekingdom}. We note that any R4 configuration beyond 72 GPUs reportedly uses optical cables across racks, and our analysis compares this configuration with an all-optical scale-up pod configuration. Best optical-vs.-electrical prefill comparisons by platform appear in Section~\ref{subsec:summary}.

\begin{table}[htbp]
\centering
\scriptsize
\setlength{\tabcolsep}{2pt}
\caption{Modeled electrical-baseline accelerators. FP8/FP4 columns are dense matmul peak PFLOPS per GPU; SU BW is unidirectional scale-up bandwidth per GPU within the electrical pod; SO BW is unidirectional scale-out (cross-rack) bandwidth per GPU modeled for devices beyond the SU pod. MoE sparse matmuls use an effective $2\times$ speedup relative to the dense compute rate. $^\dagger$R4 specs extrapolate the proposed dual-die Rubin design to a potential quad-die configuration; parameters are inferred from a roadmap report~\citep{semianalysis2025inferencekingdom} and should be read as projections.}
\label{tab:hw_all}
\resizebox{\columnwidth}{!}{%
\begin{tabular}{@{}lrrrrrrr@{}}
\toprule
\textbf{Plat.} & \textbf{FP8} & \textbf{FP4} & \textbf{Max SU} & \textbf{SU BW} & \textbf{SO BW} & \textbf{HBM} & \textbf{HBM} \\
 & \multicolumn{2}{c}{\textbf{dense (PFLOPS)}} & \textbf{GPUs} & \multicolumn{2}{c}{\textbf{uni. (GB/s)}} & \textbf{TB/s} & \textbf{GB} \\
\midrule
B200 & 4.5 & 9 & 72 & 900 & 50 & 7.67 & 192 \\
B300 & 7 & 14 & 72 & 900 & 100 & 7.94 & 288 \\
\midrule
Rubin & 17.5 & 25 & 72 & 1800 & 100 & 22 & 288 \\
R4$^\dagger$ & 35 & 50 & 576 & 1800 & 100 & 53 & 288 \\
\bottomrule
\end{tabular}%
}
\end{table}

For \emph{each} electrical baseline in Table~\ref{tab:hw_all}, we pair a matching \emph{optical} configuration: identical per-GPU compute and HBM, $4\times$ the scale-up bandwidth per GPU vs. the electrical baseline, and a maximum optical scale-up pod of 1152 GPUs (enabling the sweeps in Section~\ref{subsec:sweeps}). The $4\times$ multiplier is a conservative lower bound: Section~\ref{sec:semiconductor} notes $>$64~Tb/s of bidirectional bandwidth per GPU for 3D-integrated photonics~\citep{lightmatterpassage} vs.\ 14.4~Tb/s for current electrical NVLink~\citep{nvidiaGB200NVL72}, a ratio exceeding $4\times$. The comparison isolates interconnect bandwidth and radix; we do not model vendor-specific optical products. Appendix~\ref{sec:gb200} summarizes how Blackwell-class compute is mapped from public ISA documentation for the XLA-based cost model.

\subsection{Experimental Sweeps}
\label{subsec:sweeps}

We perform two sweeps to characterize the throughput-latency trade-off space:
\begin{enumerate}
    \item \textbf{Device sweep:} We fix the batch size per context length (8192 at 1K context length, 1024 at 8K context length, 64 at 128K context length, and 8 at 1M context length), which holds total input tokens near 8M in each case. The device count is swept over valid parallel meshes for each model: typically 8--72 GPUs for 1K and 8K input tokens (rack-scale scale-up pods in Table~\ref{tab:hw_all}), up to 288 GPUs at 128K input tokens, and up to 1152 at 1M input tokens. Prefill latency is reported vs.\ cluster size for electrical baseline and optical counterpart at each point.
    \item \textbf{Batch sweep:} We fix the device count per context length (16 at 1K context length, 72 at 8K context length, 288 at 128K context length, 1152 at 1M context length). The batch size is swept across a range. Input-token throughput and prefill latency use the same overlapped-latency pipeline as the device sweep.
\end{enumerate}

We sweep four context lengths representative of key operating regimes: 1K, 8K, 128K, and 1M tokens.

\subsection{Performance comparisons and table entries}
\label{subsec:multipliers_def}

For each evaluated configuration (model, precision, hardware, context length, sweep type), we compute \emph{overlapped} prefill latency for each point in the sweep. Overlapped prefill latency takes into account the compute--communication overlap from the XLA-based cost walk. The number reported at any given configuration (e.g., in Table~\ref{tab:sweep_heatmap}) is the ratio of overlapped latency for the electrical baseline to the overlapped latency for the optically-interconnected comparison. Values greater than $1$ denote an improvement in prefill latency when simulating optical interconnects.

For \textbf{device sweep} entries, we vary device count, for each model and GPU, and take the \textbf{maximum} multiplier across all device counts in the sweep. For \textbf{batch sweep} entries, we vary batch size at fixed device count and take the \textbf{maximum} ratio across batch sizes.

\section{Results}
\label{sec:results}

We evaluate the impact of optical interconnects on LLM inference across three workload regimes defined by context length: (i) short-context workloads (1K--8K tokens), (ii) medium-context workloads (128K tokens), and (iii) long-context workloads (1M tokens). These regimes stress different system bottlenecks—communication saturation from large batch sizes versus compute saturation leading to benefits from multi-rack scale-out—and therefore expose distinct benefits enabled by higher-bandwidth, large-radix optical interconnects.

All results in this section use the DeepSeek R1 variant (42B active parameters, FP4) on the NVIDIA B300 platform unless otherwise stated. Section~\ref{subsec:summary} summarizes results across all three models, four hardware platforms, and both sweep types for FP4 precision; Appendix~\ref{sec:full_heatmap} (Table~\ref{tab:sweep_heatmap_full}) extends this to FP8. Sections~\ref{subsec:low_context}--\ref{subsec:summary} use a \emph{bulk-prefill} sweep (fixed aggregate tokens, swept device count/batch size, up to batch 16,384) to isolate interconnect sensitivity; Section~\ref{subsec:des_check} then cross-checks these trends under the bounded per-worker batches of latency-sensitive, interactive serving.

\begin{figure*}[tbp]
\centering
\includegraphics[width=\textwidth]{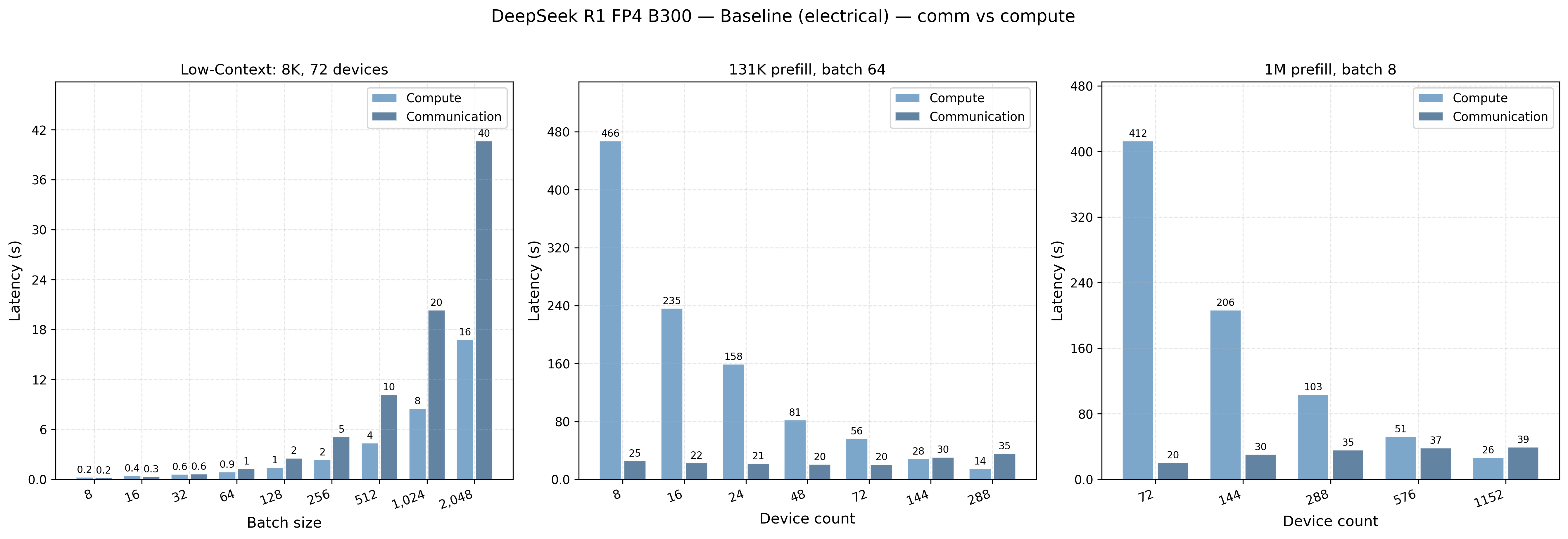}
\caption{Communication and compute component latency for representative DeepSeek R1 variant, FP4, NVIDIA B300 electrical-baseline prefill workloads --- (a) low-context 8K tokens at 72 devices across batch sizes, (b) 128K tokens at batch 64 across device counts, and (c) 1M tokens at batch 8 across device counts. The panels isolate the raw component times behind the overlapped prefill-latency curves in Figures~\ref{fig:8k_combined},~\ref{fig:131k_combined}, and~\ref{fig:1m_combined}.}
\label{fig:comm_compute_baseline}
\end{figure*}

\begin{figure*}[tbp]
\centering
\includegraphics[width=\textwidth]{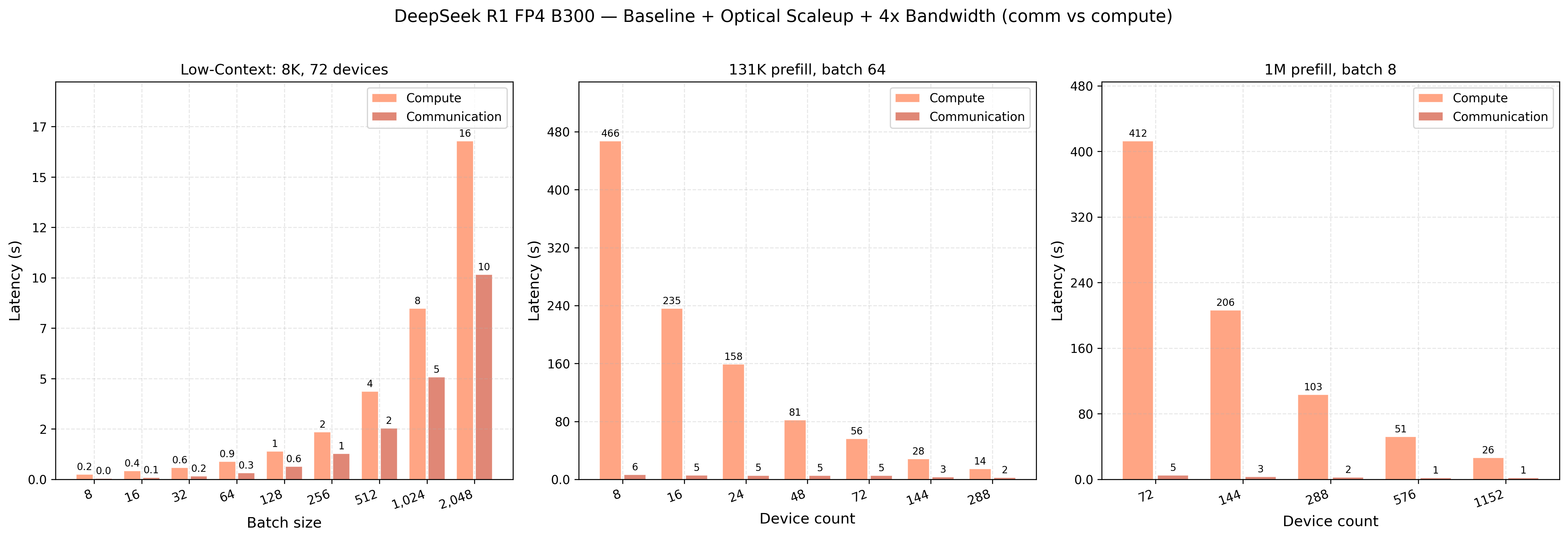}
\caption{Communication and compute component latency for the corresponding optical scale-up configuration with $4\times$ per-GPU scale-up bandwidth. Compute time is unchanged relative to the matched electrical baseline; the reduced communication component shows where high-bandwidth optical scale-up converts communication-limited cases back toward compute-limited operation.}
\label{fig:comm_compute_optical}
\end{figure*}

\subsection{Low-Context Workloads (1K and 8K Input Tokens)}
\label{subsec:low_context}

\begin{figure*}[tbp]
\centering
\includegraphics[width=\textwidth]{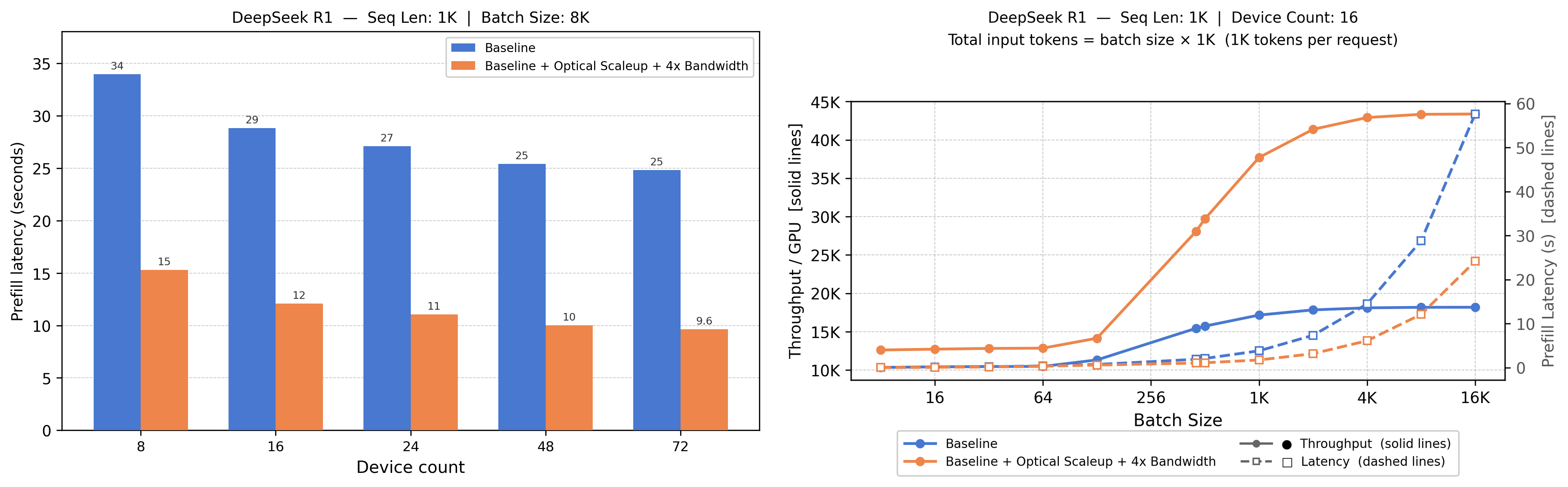}
\caption{DeepSeek R1 variant, FP4, NVIDIA B300, 1K tokens --- (a) prefill latency vs.\ number of devices, (b) input-token throughput per GPU (left axis, solid lines) and prefill latency (right axis, dashed lines) vs.\ batch size. The optical configuration (Table~\ref{tab:hw_all} pairing) achieves approximately $2.58\times$ lower latency at the best device sweep point vs.\ electrical baseline and sustains larger batch sizes at any latency SLA.}
\label{fig:1k_combined}
\end{figure*}

\begin{figure*}[tbp]
\centering
\includegraphics[width=\textwidth]{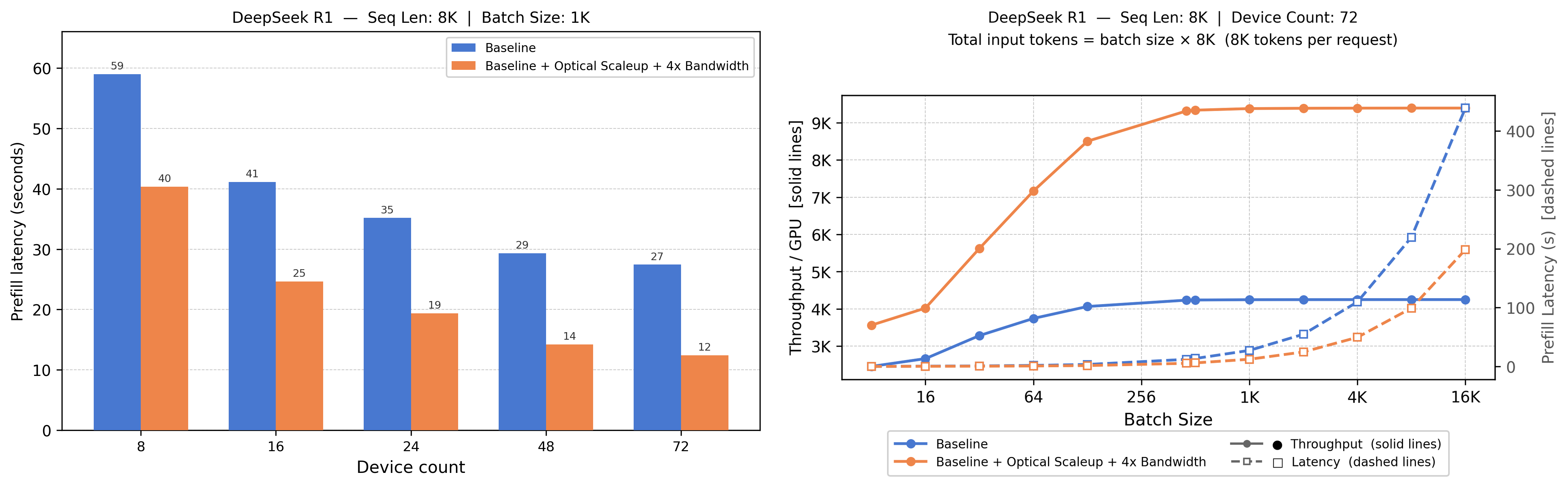}
\caption{DeepSeek R1 variant, FP4, NVIDIA B300, 8K tokens --- (a) prefill latency vs.\ number of devices, (b) input-token throughput per GPU (left axis, solid lines) and prefill latency (right axis, dashed lines) vs.\ batch size. The optical advantage in the device sweep is approximately $2.21\times$ at comparable device counts.}
\label{fig:8k_combined}
\end{figure*}

We examine short context workloads in Figures~\ref{fig:1k_combined} and~\ref{fig:8k_combined}. At short context lengths, the data reveals a counterintuitive but defining pattern: communication can dominate computation as batch size grows. As seen in panel (b) in Figures~\ref{fig:1k_combined} and~\ref{fig:8k_combined}, serving requests at very large batch sizes maximizes compute utilization per GPU. However, larger batches also create larger activation tensors that must be continuously shuffled across devices. Figure~\ref{fig:comm_compute_baseline}(a) isolates this effect for the 8K-token workload at 72 devices: by batch 2048, the electrical baseline spends approximately 40.6 seconds in communication versus 16.8 seconds in computation.

With electrical interconnects, this communication overhead becomes a significant bottleneck to throughput. As batch size increases, the volume of data exchanged in distributed collectives grows quickly enough that the electrical baseline flattens in throughput. Optical interconnects, offering $4\times$ higher bandwidth, directly reduce this communication phase: in Figure~\ref{fig:comm_compute_optical}(a), the same 8K-token, batch-2048 point spends approximately 10.2 seconds in communication while compute time remains 16.8 seconds. Consequently, the optical configuration sustains proportionally higher input-token throughput per GPU at large batches, as shown by the solid lines in panel (b) of Figures~\ref{fig:1k_combined} and~\ref{fig:8k_combined}. The dashed lines in panel (b) highlight the corresponding operational benefit: the optical configuration permits a system to meet a lower-latency SLA at larger batch sizes than the electrical infrastructure.

\subsection{Medium-Context Workloads (128K Input Tokens)}
\label{subsec:medium_context}

\begin{figure*}[tbp]
\centering
\includegraphics[width=\textwidth]{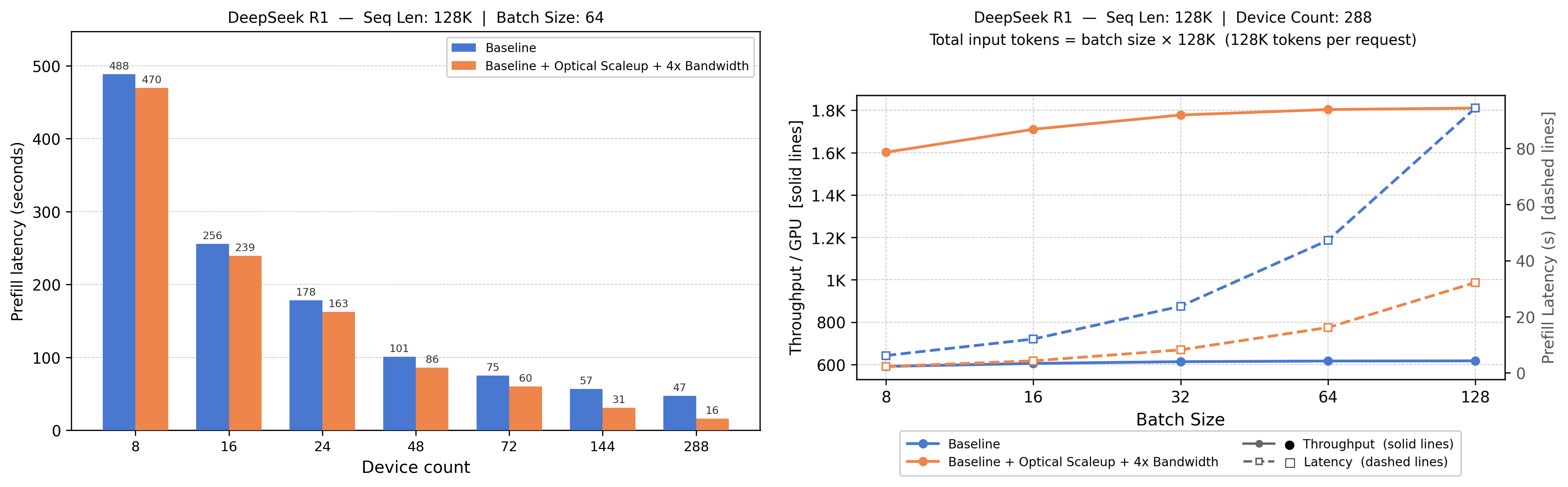}
\caption{DeepSeek R1 variant, FP4, NVIDIA B300, 128K tokens --- (a) prefill latency vs.\ number of devices, (b) input-token throughput per GPU (left axis, solid lines) and prefill latency (right axis, dashed lines) vs.\ batch size. Communication dominates at this context length; on B300 the best device-sweep factor is approximately $2.93\times$, lower than on B200/Rubin at the same aggregate electrical scale-up bandwidth because faster compute partially offsets the collective bottleneck.}
\label{fig:131k_combined}
\end{figure*}

Figure~\ref{fig:131k_combined} shows the results for 128K sequence length (288-device batch sweep).

First, on a single rack (72 GPUs), continuous inference of a 128K token sequence at batch 64 is bound by computation time. Figure~\ref{fig:comm_compute_baseline}(b) shows that at 72 devices, computation requires approximately 56.1 seconds, whereas communication takes 20.3 seconds. In order to achieve a lower TTFT SLA, an operator would be forced to scale out the workload across more devices—for instance, to 288 GPUs.

However, resolving the computational bottleneck by scaling to 288 GPUs introduces a severe interconnect penalty for electrical systems. When the device count exceeds the capacity of a single 72-device rack, cross-rack communication links become a bottleneck. In Figure~\ref{fig:comm_compute_baseline}(b), the electrical baseline at 288 devices drops compute to approximately 14.3 seconds, but communication increases to 35.5 seconds.

Optical scale-up changes the time spent doing communication versus doing computation. By introducing a homogeneous ultra-high-bandwidth scale-up pod up to 1152 devices, optical interconnects enable the compute scaling to 288 GPUs while dropping the communication time to 2.3 seconds, as shown in Figure~\ref{fig:comm_compute_optical}(b). This manifests as a $\sim$2.8--3.0$\times$ improvement at 288 devices for B300 (Table~\ref{tab:sweep_heatmap}). Rack-limited platforms like B200 and Rubin observe even larger ratios ($\sim$4.3--5.8$\times$) due to more pronounced cross-rack penalties. R4, built with a large native electro-optical scale-up pod of 576 GPUs, sees smaller but still material gains at this context length ($\sim$2.0--2.1$\times$). Panel (b) of Figure~\ref{fig:131k_combined} shows the operational consequence: high scale-up and high bandwidth are necessary for sustained batch sizes at reasonable latencies, with the solid lines showing throughput gains and the dashed lines showing the corresponding latency reductions.

\subsection{Large-Context Workloads (1M Tokens)}
\label{subsec:large_context}

\begin{figure*}[tbp]
\centering
\includegraphics[width=\textwidth]{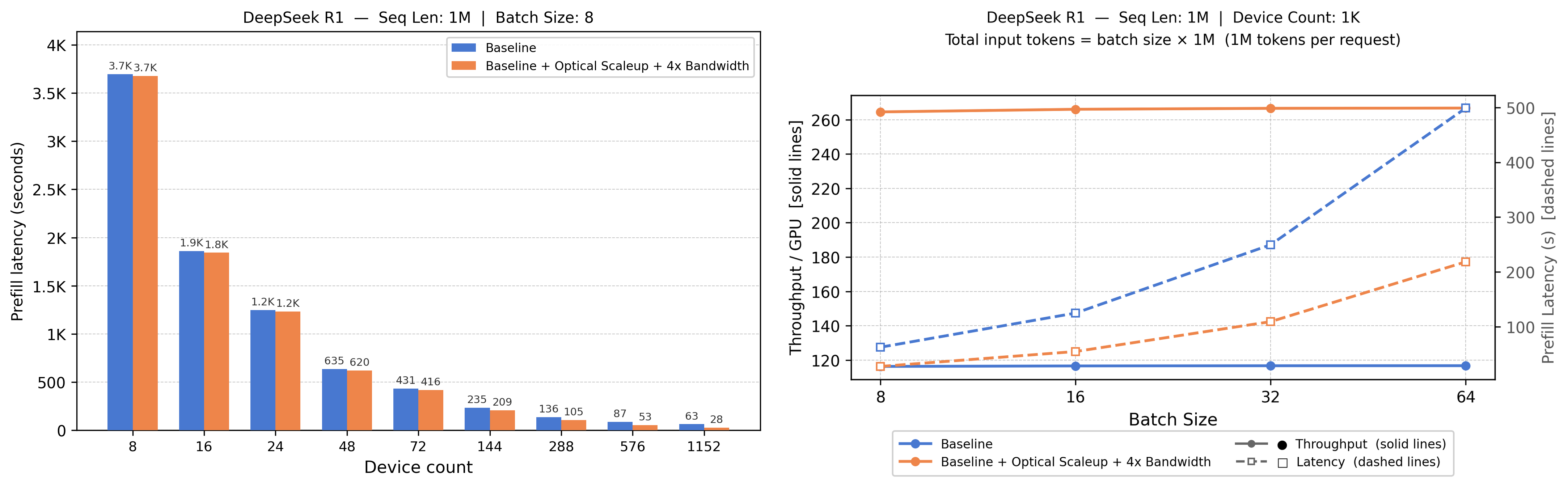}
\caption{DeepSeek R1 variant, FP4, NVIDIA B300, 1M tokens --- (a) prefill latency vs.\ number of devices, (b) input-token throughput per GPU (left axis, solid lines) and prefill latency (right axis, dashed lines) vs.\ batch size. Device sweep best-case improvement vs.\ electrical baseline is approximately $2.27\times$ on B300 at 1152 devices; B200/Rubin reach $\sim$3.4--4.5$\times$ and R4 reaches $8\times$+ in Table~\ref{tab:sweep_heatmap} because the latency floor is set more by scale-up limits than by per-GPU FLOPS alone.}
\label{fig:1m_combined}
\end{figure*}

Figure~\ref{fig:1m_combined} shows the results for a 1M sequence length over a maximum of 1152 devices. At this scale, keeping the 1152 devices in a high-bandwidth scale-up pod is essential for low prefill latency.

In panel (a) of Figure~\ref{fig:1m_combined}, single-rack (up to 72 GPUs) or lightly distributed scale-up pods are compute-time-bottlenecked. To bring processing time to lower levels for 1M tokens, the workload benefits from scaling to a higher device footprint; this analysis sweeps up to 1152 devices at batch 8. At 1152 devices, the compute time decreases to 26.2 seconds per pass, as shown in Figure~\ref{fig:comm_compute_baseline}(c).

The communication burden increases as the number of GPUs scales, and there is an inflection point when scaling the workload to multiple racks. In the Baseline systems, with up to 72 GPUs in a scale-up pod, the communication overhead partially counters the performance benefits of scaling the workload beyond 72 devices. The diminishing returns are particularly apparent when scaling from 288 to 576 devices, and again from 576 to 1152 devices. A 1152-device footprint results in 39.2 seconds of communication latency per forward pass. The higher scale-up pod and bandwidth enabled in the optical configuration (all 1152 devices in the scale-up pod) allow the workload to continue scaling nearly linearly as more devices are added, reducing communication latency to just 1.6 seconds for 1152 devices in Figure~\ref{fig:comm_compute_optical}(c). This enables an overlapping latency of 27.5 seconds, dominated by compute time, a $2.3\times$ total speedup compared to the electrically connected Baseline on the B300 (Table~\ref{tab:sweep_heatmap}). As shown in panel (b) of Figure~\ref{fig:1m_combined}, optics allow for greater throughput per GPU (solid lines) and lower prefill latency (dashed lines) as batch sizes increase. This large-context, high-batch paradigm is particularly relevant for agentic and reasoning workloads that generate many intermediate tokens per request \citep{Uptime2025,Heim2024_o1}.

\subsection{Summary Across Models, Precisions, and Hardware}
\label{subsec:summary}

Table~\ref{tab:sweep_heatmap} consolidates the multiplier sweep across all three model scales, four hardware platforms, FP4 precision, and 36 representative scenarios (Table~\ref{tab:sweep_heatmap_full} in Appendix~\ref{sec:full_heatmap} provides the corresponding FP8 numbers). Each cell is the best electrical-to-optical overlapped prefill-latency ratio for that configuration. The header rows list the sequence length, total-token tier, device count, and batch size for each scenario. The total-token tier is important because it fixes the derived batch size at each sequence length; consequently, two columns with the same context length can expose different communication pressure if they represent different aggregate input-token volumes.

\begin{table*}[!t]
\centering
\caption{Sweep best multiplier heatmap (FP4 quantization). Device columns use \texttt{dev} $\in$ \{16, 72, 288, 576, 1152\} (capped by \texttt{seq\_len}); the underlying sweep includes more device counts in \texttt{summary.csv}. Three total-token tiers per column (multiples of $1024^2$; batch $=$ tokens$/$\texttt{seq\_len}, capped). Tiers depend on \texttt{dev}: \texttt{[8,72]} $\rightarrow$ \texttt{2M}/\texttt{8M}/\texttt{16M}; \texttt{(72,288]} $\rightarrow$ \texttt{2M}/\texttt{8M}/\texttt{32M}; \texttt{(288,1152]} $\rightarrow$ \texttt{8M}/\texttt{32M}/\texttt{128M}. Header \texttt{seq\_len} uses binary-style \texttt{1K}${}=1024$, \texttt{8K}${}=8{\times}1024$, \texttt{128K}${}=128{\times}1024$, \texttt{1M}${}=1024^2$. Cell color scales from min to max among these heatmap cells only. For FP8 results see Table~\ref{tab:sweep_heatmap_full} in Appendix~\ref{sec:full_heatmap}.}
\label{tab:sweep_heatmap}
\scriptsize
\setlength{\tabcolsep}{2pt}
\renewcommand{\arraystretch}{1.28}
\resizebox{\textwidth}{!}{%
\begin{tabular}{llll|rrrrrrrrrrrrrrrrrrrrrrrrrrrrrrrrrrrr}
\toprule
Model & Hardware & Quant & & \multicolumn{36}{c}{Best multiplier (optical vs.\ no-optical)} \\
\cmidrule(lr){5-40}
 & & & seq\_len & \texttt{1K} & \texttt{1K} & \texttt{1K} & \texttt{1K} & \texttt{1K} & \texttt{1K} & \texttt{8K} & \texttt{8K} & \texttt{8K} & \texttt{8K} & \texttt{8K} & \texttt{8K} & \texttt{128K} & \texttt{128K} & \texttt{128K} & \texttt{128K} & \texttt{128K} & \texttt{128K} & \texttt{128K} & \texttt{128K} & \texttt{128K} & \texttt{1M} & \texttt{1M} & \texttt{1M} & \texttt{1M} & \texttt{1M} & \texttt{1M} & \texttt{1M} & \texttt{1M} & \texttt{1M} & \texttt{1M} & \texttt{1M} & \texttt{1M} & \texttt{1M} & \texttt{1M} & \texttt{1M} \\
 & & & tokens & \texttt{2M} & \texttt{8M} & \texttt{16M} & \texttt{2M} & \texttt{8M} & \texttt{16M} & \texttt{2M} & \texttt{8M} & \texttt{16M} & \texttt{2M} & \texttt{8M} & \texttt{16M} & \texttt{2M} & \texttt{8M} & \texttt{16M} & \texttt{2M} & \texttt{8M} & \texttt{16M} & \texttt{2M} & \texttt{8M} & \texttt{32M} & \texttt{2M} & \texttt{8M} & \texttt{16M} & \texttt{2M} & \texttt{8M} & \texttt{16M} & \texttt{2M} & \texttt{8M} & \texttt{32M} & \texttt{8M} & \texttt{32M} & \texttt{128M} & \texttt{8M} & \texttt{32M} & \texttt{128M} \\
 & & & dev & 16 & 16 & 16 & 72 & 72 & 72 & 16 & 16 & 16 & 72 & 72 & 72 & 16 & 16 & 16 & 72 & 72 & 72 & 288 & 288 & 288 & 16 & 16 & 16 & 72 & 72 & 72 & 288 & 288 & 288 & 576 & 576 & 576 & 1152 & 1152 & 1152 \\
 & & & batch & 2048 & 8192 & 16384 & 2048 & 8192 & 16384 & 256 & 1024 & 2048 & 256 & 1024 & 2048 & 16 & 64 & 128 & 16 & 64 & 128 & 16 & 64 & 256 & 2 & 8 & 16 & 2 & 8 & 16 & 2 & 8 & 32 & 8 & 32 & 64 & 8 & 32 & 64 \\
\midrule
mini & B200 & FP4 & & \cellcolor{heatblue!24}2.12 & \cellcolor{heatblue!25}2.23 & \cellcolor{heatblue!25}2.24 & \cellcolor{heatblue!26}2.33 & \cellcolor{heatblue!27}2.41 & \cellcolor{heatblue!27}2.41 & \cellcolor{heatblue!18}1.59 & \cellcolor{heatblue!19}1.61 & \cellcolor{heatblue!19}1.62 & \cellcolor{heatblue!23}2.04 & \cellcolor{heatblue!24}2.09 & \cellcolor{heatblue!24}2.09 & \cellcolor{heatblue!13}1.06 & \cellcolor{heatblue!13}1.07 & \cellcolor{heatblue!13}1.07 & \cellcolor{heatblue!14}1.23 & \cellcolor{heatblue!14}1.23 & \cellcolor{heatblue!14}1.23 & \cellcolor{heatblue!44}3.98 & \cellcolor{heatblue!47}4.27 & \cellcolor{heatblue!47}4.30 & \cellcolor{heatblue!12}1.01 & \cellcolor{heatblue!12}1.01 & \cellcolor{heatblue!12}1.01 & \cellcolor{heatblue!12}1.03 & \cellcolor{heatblue!12}1.03 & \cellcolor{heatblue!12}1.03 & \cellcolor{heatblue!17}1.50 & \cellcolor{heatblue!17}1.51 & \cellcolor{heatblue!17}1.51 & \cellcolor{heatblue!24}2.12 & \cellcolor{heatblue!24}2.13 & \cellcolor{heatblue!24}2.13 & \cellcolor{heatblue!36}3.29 & \cellcolor{heatblue!37}3.32 & \cellcolor{heatblue!37}3.32 \\
\addlinespace[2pt]
 & B300 & FP4 & & \cellcolor{heatblue!26}2.34 & \cellcolor{heatblue!27}2.41 & \cellcolor{heatblue!27}2.41 & \cellcolor{heatblue!29}2.59 & \cellcolor{heatblue!29}2.64 & \cellcolor{heatblue!30}2.64 & \cellcolor{heatblue!19}1.65 & \cellcolor{heatblue!19}1.67 & \cellcolor{heatblue!19}1.67 & \cellcolor{heatblue!25}2.21 & \cellcolor{heatblue!25}2.24 & \cellcolor{heatblue!25}2.24 & \cellcolor{heatblue!13}1.07 & \cellcolor{heatblue!13}1.07 & \cellcolor{heatblue!13}1.07 & \cellcolor{heatblue!15}1.25 & \cellcolor{heatblue!15}1.25 & \cellcolor{heatblue!15}1.25 & \cellcolor{heatblue!31}2.75 & \cellcolor{heatblue!32}2.84 & \cellcolor{heatblue!32}2.85 & \cellcolor{heatblue!12}1.01 & \cellcolor{heatblue!12}1.01 & \cellcolor{heatblue!12}1.01 & \cellcolor{heatblue!12}1.04 & \cellcolor{heatblue!12}1.04 & \cellcolor{heatblue!12}1.04 & \cellcolor{heatblue!15}1.28 & \cellcolor{heatblue!15}1.28 & \cellcolor{heatblue!15}1.28 & \cellcolor{heatblue!18}1.60 & \cellcolor{heatblue!18}1.60 & \cellcolor{heatblue!18}1.60 & \cellcolor{heatblue!25}2.20 & \cellcolor{heatblue!25}2.21 & \cellcolor{heatblue!25}2.21 \\
\addlinespace[2pt]
 & Rubin & FP4 & & \cellcolor{heatblue!28}2.48 & \cellcolor{heatblue!29}2.62 & \cellcolor{heatblue!29}2.62 & \cellcolor{heatblue!30}2.71 & \cellcolor{heatblue!31}2.80 & \cellcolor{heatblue!31}2.81 & \cellcolor{heatblue!21}1.81 & \cellcolor{heatblue!21}1.85 & \cellcolor{heatblue!21}1.85 & \cellcolor{heatblue!27}2.38 & \cellcolor{heatblue!27}2.45 & \cellcolor{heatblue!27}2.45 & \cellcolor{heatblue!13}1.10 & \cellcolor{heatblue!13}1.10 & \cellcolor{heatblue!13}1.10 & \cellcolor{heatblue!15}1.33 & \cellcolor{heatblue!16}1.34 & \cellcolor{heatblue!16}1.34 & \cellcolor{heatblue!57}5.16 & \cellcolor{heatblue!60}5.51 & \cellcolor{heatblue!61}5.55 & \cellcolor{heatblue!12}1.01 & \cellcolor{heatblue!12}1.01 & \cellcolor{heatblue!12}1.01 & \cellcolor{heatblue!12}1.05 & \cellcolor{heatblue!12}1.05 & \cellcolor{heatblue!12}1.05 & \cellcolor{heatblue!20}1.71 & \cellcolor{heatblue!20}1.72 & \cellcolor{heatblue!20}1.73 & \cellcolor{heatblue!29}2.59 & \cellcolor{heatblue!29}2.60 & \cellcolor{heatblue!29}2.60 & \cellcolor{heatblue!47}4.24 & \cellcolor{heatblue!47}4.28 & \cellcolor{heatblue!47}4.28 \\
\addlinespace[2pt]
 & R4 & FP4 & & \cellcolor{heatblue!33}3.00 & \cellcolor{heatblue!34}3.10 & \cellcolor{heatblue!35}3.11 & \cellcolor{heatblue!35}3.11 & \cellcolor{heatblue!35}3.18 & \cellcolor{heatblue!35}3.19 & \cellcolor{heatblue!26}2.36 & \cellcolor{heatblue!27}2.41 & \cellcolor{heatblue!27}2.42 & \cellcolor{heatblue!33}2.92 & \cellcolor{heatblue!33}2.98 & \cellcolor{heatblue!33}2.99 & \cellcolor{heatblue!14}1.21 & \cellcolor{heatblue!14}1.21 & \cellcolor{heatblue!14}1.22 & \cellcolor{heatblue!19}1.66 & \cellcolor{heatblue!19}1.67 & \cellcolor{heatblue!19}1.67 & \cellcolor{heatblue!22}1.91 & \cellcolor{heatblue!22}1.98 & \cellcolor{heatblue!23}1.99 & \cellcolor{heatblue!12}1.03 & \cellcolor{heatblue!12}1.03 & \cellcolor{heatblue!12}1.03 & \cellcolor{heatblue!13}1.11 & \cellcolor{heatblue!13}1.11 & \cellcolor{heatblue!13}1.11 & \cellcolor{heatblue!14}1.19 & \cellcolor{heatblue!14}1.19 & \cellcolor{heatblue!14}1.19 & \cellcolor{heatblue!15}1.28 & \cellcolor{heatblue!15}1.28 & \cellcolor{heatblue!15}1.28 & \cellcolor{heatblue!87}8.04 & \cellcolor{heatblue!89}8.14 & \cellcolor{heatblue!89}8.15 \\
\midrule
r1 & B200 & FP4 & & \cellcolor{heatblue!24}2.11 & \cellcolor{heatblue!25}2.21 & \cellcolor{heatblue!25}2.22 & \cellcolor{heatblue!26}2.28 & \cellcolor{heatblue!26}2.35 & \cellcolor{heatblue!26}2.36 & \cellcolor{heatblue!18}1.59 & \cellcolor{heatblue!19}1.62 & \cellcolor{heatblue!19}1.62 & \cellcolor{heatblue!23}2.01 & \cellcolor{heatblue!23}2.06 & \cellcolor{heatblue!23}2.06 & \cellcolor{heatblue!13}1.07 & \cellcolor{heatblue!13}1.07 & \cellcolor{heatblue!13}1.07 & \cellcolor{heatblue!14}1.23 & \cellcolor{heatblue!14}1.23 & \cellcolor{heatblue!14}1.23 & \cellcolor{heatblue!45}4.12 & \cellcolor{heatblue!49}4.42 & \cellcolor{heatblue!49}4.46 & \cellcolor{heatblue!12}1.01 & \cellcolor{heatblue!12}1.01 & \cellcolor{heatblue!12}1.01 & \cellcolor{heatblue!12}1.03 & \cellcolor{heatblue!12}1.03 & \cellcolor{heatblue!12}1.03 & \cellcolor{heatblue!18}1.53 & \cellcolor{heatblue!18}1.53 & \cellcolor{heatblue!18}1.54 & \cellcolor{heatblue!25}2.18 & \cellcolor{heatblue!25}2.19 & \cellcolor{heatblue!25}2.19 & \cellcolor{heatblue!38}3.41 & \cellcolor{heatblue!38}3.45 & \cellcolor{heatblue!38}3.45 \\
\addlinespace[2pt]
 & B300 & FP4 & & \cellcolor{heatblue!26}2.32 & \cellcolor{heatblue!27}2.38 & \cellcolor{heatblue!27}2.38 & \cellcolor{heatblue!28}2.53 & \cellcolor{heatblue!29}2.58 & \cellcolor{heatblue!29}2.58 & \cellcolor{heatblue!19}1.65 & \cellcolor{heatblue!19}1.67 & \cellcolor{heatblue!19}1.67 & \cellcolor{heatblue!25}2.18 & \cellcolor{heatblue!25}2.21 & \cellcolor{heatblue!25}2.21 & \cellcolor{heatblue!13}1.07 & \cellcolor{heatblue!13}1.07 & \cellcolor{heatblue!13}1.07 & \cellcolor{heatblue!15}1.24 & \cellcolor{heatblue!15}1.25 & \cellcolor{heatblue!15}1.25 & \cellcolor{heatblue!32}2.83 & \cellcolor{heatblue!33}2.93 & \cellcolor{heatblue!33}2.93 & \cellcolor{heatblue!12}1.01 & \cellcolor{heatblue!12}1.01 & \cellcolor{heatblue!12}1.01 & \cellcolor{heatblue!12}1.04 & \cellcolor{heatblue!12}1.04 & \cellcolor{heatblue!12}1.04 & \cellcolor{heatblue!15}1.29 & \cellcolor{heatblue!15}1.30 & \cellcolor{heatblue!15}1.30 & \cellcolor{heatblue!19}1.63 & \cellcolor{heatblue!19}1.63 & \cellcolor{heatblue!19}1.63 & \cellcolor{heatblue!26}2.27 & \cellcolor{heatblue!26}2.28 & \cellcolor{heatblue!26}2.28 \\
\addlinespace[2pt]
 & Rubin & FP4 & & \cellcolor{heatblue!28}2.45 & \cellcolor{heatblue!29}2.58 & \cellcolor{heatblue!29}2.58 & \cellcolor{heatblue!30}2.64 & \cellcolor{heatblue!30}2.72 & \cellcolor{heatblue!30}2.73 & \cellcolor{heatblue!21}1.81 & \cellcolor{heatblue!21}1.85 & \cellcolor{heatblue!21}1.85 & \cellcolor{heatblue!26}2.34 & \cellcolor{heatblue!27}2.40 & \cellcolor{heatblue!27}2.40 & \cellcolor{heatblue!13}1.10 & \cellcolor{heatblue!13}1.10 & \cellcolor{heatblue!13}1.10 & \cellcolor{heatblue!15}1.33 & \cellcolor{heatblue!15}1.33 & \cellcolor{heatblue!15}1.33 & \cellcolor{heatblue!58}5.34 & \cellcolor{heatblue!62}5.71 & \cellcolor{heatblue!63}5.75 & \cellcolor{heatblue!12}1.01 & \cellcolor{heatblue!12}1.01 & \cellcolor{heatblue!12}1.01 & \cellcolor{heatblue!12}1.05 & \cellcolor{heatblue!12}1.05 & \cellcolor{heatblue!12}1.05 & \cellcolor{heatblue!20}1.75 & \cellcolor{heatblue!20}1.76 & \cellcolor{heatblue!20}1.76 & \cellcolor{heatblue!30}2.68 & \cellcolor{heatblue!30}2.69 & \cellcolor{heatblue!30}2.69 & \cellcolor{heatblue!48}4.41 & \cellcolor{heatblue!49}4.45 & \cellcolor{heatblue!49}4.45 \\
\addlinespace[2pt]
 & R4 & FP4 & & \cellcolor{heatblue!33}2.95 & \cellcolor{heatblue!34}3.05 & \cellcolor{heatblue!34}3.06 & \cellcolor{heatblue!34}3.05 & \cellcolor{heatblue!35}3.13 & \cellcolor{heatblue!35}3.13 & \cellcolor{heatblue!26}2.34 & \cellcolor{heatblue!27}2.39 & \cellcolor{heatblue!27}2.40 & \cellcolor{heatblue!32}2.86 & \cellcolor{heatblue!32}2.92 & \cellcolor{heatblue!33}2.92 & \cellcolor{heatblue!14}1.21 & \cellcolor{heatblue!14}1.22 & \cellcolor{heatblue!14}1.22 & \cellcolor{heatblue!19}1.66 & \cellcolor{heatblue!19}1.67 & \cellcolor{heatblue!19}1.67 & \cellcolor{heatblue!22}1.94 & \cellcolor{heatblue!23}2.01 & \cellcolor{heatblue!23}2.02 & \cellcolor{heatblue!12}1.03 & \cellcolor{heatblue!12}1.03 & \cellcolor{heatblue!12}1.03 & \cellcolor{heatblue!13}1.11 & \cellcolor{heatblue!13}1.11 & \cellcolor{heatblue!13}1.11 & \cellcolor{heatblue!14}1.20 & \cellcolor{heatblue!14}1.20 & \cellcolor{heatblue!14}1.20 & \cellcolor{heatblue!15}1.31 & \cellcolor{heatblue!15}1.31 & \cellcolor{heatblue!15}1.31 & \cellcolor{heatblue!91}8.33 & \cellcolor{heatblue!92}8.43 & \cellcolor{heatblue!92}8.44 \\
\midrule
next & B200 & FP4 & & \cellcolor{heatblue!25}2.18 & \cellcolor{heatblue!25}2.23 & \cellcolor{heatblue!25}2.23 & \cellcolor{heatblue!26}2.33 & \cellcolor{heatblue!26}2.36 & \cellcolor{heatblue!27}2.36 & \cellcolor{heatblue!19}1.62 & \cellcolor{heatblue!19}1.63 & \cellcolor{heatblue!19}1.63 & \cellcolor{heatblue!23}2.05 & \cellcolor{heatblue!23}2.07 & \cellcolor{heatblue!23}2.08 & \cellcolor{heatblue!13}1.07 & \cellcolor{heatblue!13}1.07 & \cellcolor{heatblue!13}1.07 & \cellcolor{heatblue!14}1.24 & \cellcolor{heatblue!14}1.24 & \cellcolor{heatblue!14}1.24 & \cellcolor{heatblue!48}4.38 & \cellcolor{heatblue!50}4.52 & \cellcolor{heatblue!50}4.53 & \cellcolor{heatblue!12}1.01 & \cellcolor{heatblue!12}1.01 & \cellcolor{heatblue!12}1.01 & \cellcolor{heatblue!12}1.03 & \cellcolor{heatblue!12}1.03 & \cellcolor{heatblue!12}1.03 & \cellcolor{heatblue!18}1.55 & \cellcolor{heatblue!18}1.55 & \cellcolor{heatblue!18}1.55 & \cellcolor{heatblue!25}2.22 & \cellcolor{heatblue!25}2.23 & \cellcolor{heatblue!25}2.23 & \cellcolor{heatblue!39}3.50 & \cellcolor{heatblue!39}3.51 & \cellcolor{heatblue!39}3.51 \\
\addlinespace[2pt]
 & B300 & FP4 & & \cellcolor{heatblue!27}2.38 & \cellcolor{heatblue!27}2.40 & \cellcolor{heatblue!27}2.40 & \cellcolor{heatblue!29}2.57 & \cellcolor{heatblue!29}2.59 & \cellcolor{heatblue!29}2.59 & \cellcolor{heatblue!19}1.68 & \cellcolor{heatblue!19}1.68 & \cellcolor{heatblue!19}1.68 & \cellcolor{heatblue!25}2.21 & \cellcolor{heatblue!25}2.23 & \cellcolor{heatblue!25}2.23 & \cellcolor{heatblue!13}1.07 & \cellcolor{heatblue!13}1.07 & \cellcolor{heatblue!13}1.07 & \cellcolor{heatblue!15}1.25 & \cellcolor{heatblue!15}1.25 & \cellcolor{heatblue!15}1.25 & \cellcolor{heatblue!33}2.95 & \cellcolor{heatblue!33}2.99 & \cellcolor{heatblue!33}2.99 & \cellcolor{heatblue!12}1.01 & \cellcolor{heatblue!12}1.01 & \cellcolor{heatblue!12}1.01 & \cellcolor{heatblue!12}1.04 & \cellcolor{heatblue!12}1.04 & \cellcolor{heatblue!12}1.04 & \cellcolor{heatblue!15}1.31 & \cellcolor{heatblue!15}1.31 & \cellcolor{heatblue!15}1.31 & \cellcolor{heatblue!19}1.66 & \cellcolor{heatblue!19}1.66 & \cellcolor{heatblue!19}1.66 & \cellcolor{heatblue!26}2.33 & \cellcolor{heatblue!26}2.33 & \cellcolor{heatblue!26}2.33 \\
\addlinespace[2pt]
 & Rubin & FP4 & & \cellcolor{heatblue!28}2.53 & \cellcolor{heatblue!29}2.58 & \cellcolor{heatblue!29}2.59 & \cellcolor{heatblue!30}2.69 & \cellcolor{heatblue!30}2.73 & \cellcolor{heatblue!31}2.73 & \cellcolor{heatblue!21}1.85 & \cellcolor{heatblue!21}1.86 & \cellcolor{heatblue!21}1.86 & \cellcolor{heatblue!27}2.38 & \cellcolor{heatblue!27}2.41 & \cellcolor{heatblue!27}2.41 & \cellcolor{heatblue!13}1.10 & \cellcolor{heatblue!13}1.10 & \cellcolor{heatblue!13}1.10 & \cellcolor{heatblue!16}1.34 & \cellcolor{heatblue!16}1.34 & \cellcolor{heatblue!16}1.34 & \cellcolor{heatblue!62}5.67 & \cellcolor{heatblue!64}5.82 & \cellcolor{heatblue!64}5.84 & \cellcolor{heatblue!12}1.01 & \cellcolor{heatblue!12}1.01 & \cellcolor{heatblue!12}1.01 & \cellcolor{heatblue!12}1.05 & \cellcolor{heatblue!12}1.05 & \cellcolor{heatblue!12}1.05 & \cellcolor{heatblue!20}1.79 & \cellcolor{heatblue!20}1.79 & \cellcolor{heatblue!20}1.79 & \cellcolor{heatblue!31}2.74 & \cellcolor{heatblue!31}2.74 & \cellcolor{heatblue!31}2.74 & \cellcolor{heatblue!50}4.51 & \cellcolor{heatblue!50}4.53 & \cellcolor{heatblue!50}4.53 \\
\addlinespace[2pt]
 & R4 & FP4 & & \cellcolor{heatblue!33}3.01 & \cellcolor{heatblue!34}3.05 & \cellcolor{heatblue!34}3.05 & \cellcolor{heatblue!34}3.10 & \cellcolor{heatblue!35}3.13 & \cellcolor{heatblue!35}3.13 & \cellcolor{heatblue!27}2.38 & \cellcolor{heatblue!27}2.40 & \cellcolor{heatblue!27}2.40 & \cellcolor{heatblue!32}2.89 & \cellcolor{heatblue!32}2.91 & \cellcolor{heatblue!32}2.92 & \cellcolor{heatblue!14}1.22 & \cellcolor{heatblue!14}1.22 & \cellcolor{heatblue!14}1.22 & \cellcolor{heatblue!19}1.67 & \cellcolor{heatblue!19}1.68 & \cellcolor{heatblue!19}1.68 & \cellcolor{heatblue!23}2.06 & \cellcolor{heatblue!24}2.09 & \cellcolor{heatblue!24}2.10 & \cellcolor{heatblue!12}1.03 & \cellcolor{heatblue!12}1.03 & \cellcolor{heatblue!12}1.03 & \cellcolor{heatblue!13}1.12 & \cellcolor{heatblue!13}1.12 & \cellcolor{heatblue!13}1.12 & \cellcolor{heatblue!14}1.22 & \cellcolor{heatblue!14}1.22 & \cellcolor{heatblue!14}1.22 & \cellcolor{heatblue!16}1.35 & \cellcolor{heatblue!16}1.35 & \cellcolor{heatblue!16}1.35 & \cellcolor{heatblue!92}8.42 & \cellcolor{heatblue!92}8.46 & \cellcolor{heatblue!92}8.47 \\
\bottomrule
\end{tabular}
}
\end{table*}

Several consistent patterns emerge from this analysis:

\paragraph{GPU type and electrical scale-up pod shape the performance benefits of implementing optical interconnects}
At 1K--8K context length, the performance benefits of optical interconnects generally increase from B200 to B300 to Rubin to R4 (Table~\ref{tab:sweep_heatmap}): faster GPUs shrink the compute component of prefill, so the same collective traffic accounts for a larger fraction of latency. FP4 shows larger multipliers than FP8 for the same reason---faster arithmetic exposes communication more---a pattern visible across all platforms in Table~\ref{tab:sweep_heatmap_full}; this reflects the compute/communication balance, not a change in communication volume.

\paragraph{Context length, total tokens, and the compute--communication balance}
Moving from 1K to 8K at the same total-token tier reduces the derived batch size eightfold and increases per-sequence computation, so the multiplier can dip even though the context is longer, since communication is less dominant---this is why 8K gains are often lower than 1K gains. Conversely, the low-gain cells around 128K at 16--72 devices and 1M at 16--288 devices have not yet crossed the most expensive scale-up boundary or remain compute-dominated: optical interconnects do not provide a uniform speedup when the baseline is not communication-limited.

\paragraph{Device count and scale-up-pod transitions}
The largest changes occur when the device count needed to reduce compute time exceeds the native electrical scale-up pod. At 128K, the 288-device columns are substantially larger than the 16- and 72-device columns for rack-limited platforms, since the workload has crossed from intra-rack scale-up into slower scale-out communication; the same mechanism appears at larger scale at 1M, where electrical baselines pay a large penalty once the footprint spills beyond the high-bandwidth pod. Because the optical configuration improves both bandwidth and radix, the observed speedup can exceed what a simple $4\times$ bandwidth argument alone would predict.

\paragraph{Hardware generation and scale-out bandwidth}
The B200-to-B300 comparison at 128K and 288 devices reflects two countervailing effects: B300's faster compute would tend to raise the value of communication bandwidth, but its higher modeled scale-out bandwidth modestly reduces the electrical penalty at the 288-device boundary, so B300 shows slightly smaller optical multipliers than compute scaling alone would predict. The relevant variable is thus not per-GPU FLOPS alone, but the ratio among compute rate, scale-up bandwidth, scale-out bandwidth, and required device count.

\paragraph{Rubin and R4}
Rubin illustrates the cost of pairing very high per-GPU compute with a 72-GPU electrical scale-up pod: at 128K and 288 devices, its rack-limited baseline incurs a large scale-out communication cost, producing the largest medium-context multipliers ($\sim$5.5--5.8$\times$). R4's 576-GPU hybrid pod already covers this regime, so its optical gains there are smaller ($\sim$2.0--2.1$\times$); but this larger native pod is insufficient at 1M, 1152 devices, where the all-optical configuration again avoids the scale-out penalty and produces the largest multipliers in the table ($\sim$8.0--8.5$\times$). This shows large scale-up pods matter most for long-context workloads that force device footprints beyond a single rack.

\paragraph{Model scale}
Across Mini, R1, and Next, the table's qualitative structure is stable: gains are modest in compute-bound cells, increase with faster arithmetic at low context, and peak when long-context workloads exceed the native scale-up pod. The exact multiplier varies with model size, but the dominant predictor remains whether collective communication lies on the overlapped-latency critical path.

\vspace{-4pt}
\subsection{Serving-Level DES Validation}
\label{subsec:des_check}

\begin{figure*}[tbp]
\centering
\vspace{-4pt}
\includegraphics[width=\textwidth,height=0.19\textheight,keepaspectratio]{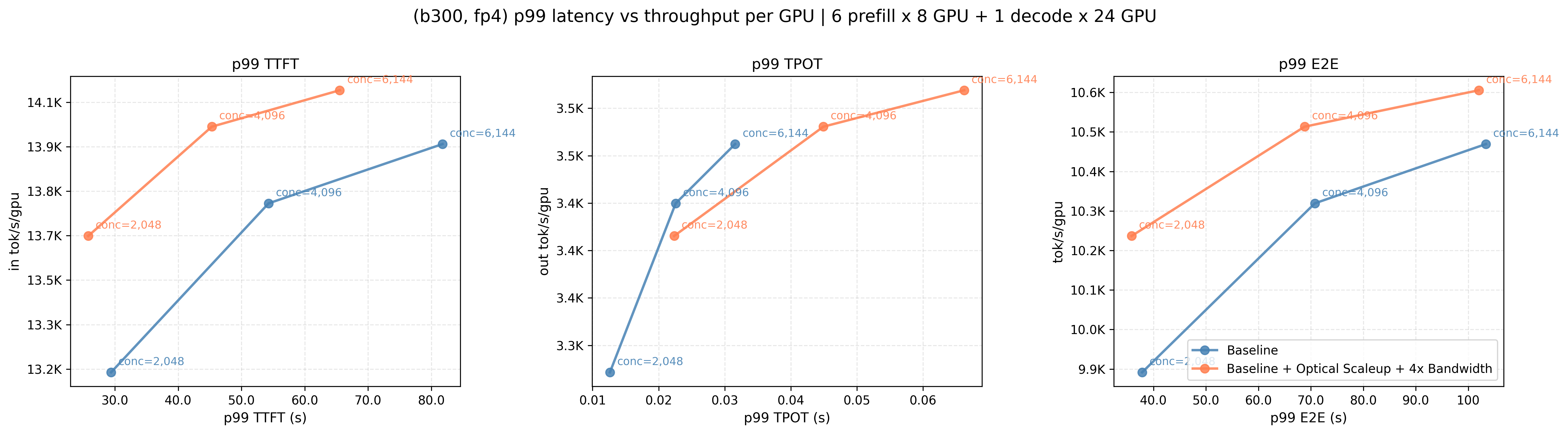}
\vspace{-6pt}
\caption{DeepSeek R1 variant, B300 FP4, disaggregated serving DES --- p99 operating points for a 72-GPU layout (six 8-GPU prefill workers, one 24-GPU decode worker), Poisson arrivals at $\lambda{=}250$~req/s, ISL~8192, OSL~1024. Each column is one SLO axis (TTFT, TPOT, end-to-end latency); each datapoint is annotated with its in-flight concurrency (2048, 4096, 6144). Blue: electrical baseline. Coral: optical scale-up ($4\times$ per-GPU bandwidth).}
\label{fig:des_p99_b300_fp4}
\end{figure*}

The analytical sweep examines bulk-prefill regimes at total token volumes up to 128M, where batch sizes can reach 16,384 at short contexts---realistic for offline, throughput-oriented workloads (e.g., prompt-cache warming, document indexing) but not for interactive chat or agentic serving. In the latter, no system waits for batches that large to accumulate before dispatching: each prefill worker must process requests promptly to meet TTFT SLOs. To quantify how optical scale-up behaves in this complementary small-batch regime, we run a discrete-event simulation (DES) on the same B300 FP4 platform. Six 8-GPU prefill workers operate at \texttt{max\_prefill\_batch\_size}~=~64 ($\sim$524K tokens per dispatch at input sequence length (ISL)~8192, roughly the low end of the bulk-sweep range). We deliberately provision a single 24-GPU decode worker to isolate how prefill-side bandwidth improvements propagate to---and interact with---the decode stage. Requests arrive at Poisson($\lambda{=}250$~req/s) across three in-flight concurrencies (2048, 4096, 6144). Figure~\ref{fig:des_p99_b300_fp4} plots the p99 operating points; Appendix~\ref{sec:des_appendix} provides the full mean/median/p99 breakdown.

Even at these modest per-worker batch sizes, optical scale-up produces compounding TTFT benefits. Shorter per-step prefill time reduces the execution latency for each dispatch; because the six workers are nearly fully utilized at high concurrency (mean batch $\approx$50~of~64), the prefill queue also drains faster, compressing queue-wait in tandem. Together, these two effects reduce p99 TTFT by 12--20\% across all concurrencies (29.4~s $\to$ 25.8~s at concurrency~2048; 81.8~s $\to$ 65.5~s at concurrency~6144). Input-token throughput improves by 1--4\% across the same range (13,959 $\to$ 14,141~tokens/s/GPU at concurrency~6144), confirming that the prefill workers were already close to their rate ceiling under both configurations.

The intentionally single decode worker exposes the downstream consequence. The 24-GPU worker is fully utilized at every concurrency tested (zero scheduler-induced idle time with a non-empty queue). Faster prefill delivers completed requests to the decode queue sooner and in tighter clusters, pushing the steady-state decode batch toward its cap ($\sim$694 $\to$ 734 at concurrency~6144). A larger decode batch produces more tokens per iteration but takes proportionally longer per token---the canonical TPOT inflation of a memory-bandwidth-bound worker running at higher load. Consequently, p99 TPOT increases by 77--110\% (31.6~ms $\to$ 66.3~ms at concurrency~6144). Because TPOT inflation and TTFT improvement act on different phases of end-to-end latency, their interaction at the tail depends on output length: for output sequence length (OSL)~1024, the two effects largely cancel, leaving p99 end-to-end latency approximately neutral ($-1$ to $-5\%$; 103.4~s $\to$ 102.0~s at concurrency~6144). Per-GPU output throughput increases only 1--4\% (3490 $\to$ 3535~tokens/s/GPU at concurrency~6144), confirming that widening the decode stage---not the prefill side---is the lever required to grow total output capacity.

Read alongside Table~\ref{tab:sweep_heatmap}, these results clarify the scope of the prefill multipliers: optical scale-up is a TTFT and input-throughput lever whose end-to-end impact is shaped by the decode-side configuration. Operators sizing a disaggregated system should treat the prefill multipliers as available headroom to reduce TTFT, and must co-design decode capacity to determine how much of that headroom converts into end-to-end latency or throughput improvements.

\vspace{-8pt}
\section{Related Work}
\label{sec:related}

AI infrastructure today combines tightly coupled electrical scale-up fabrics such as NVLink/NVSwitch and Infinity Fabric with scale-out networks such as InfiniBand, but these links face worsening power, reach, and signal-integrity limits as data rates rise \citep{nvidia2025rubin,Tokenring2026}. Compute Express Link (CXL) and Universal Chiplet Interconnect Express (UCIe) address adjacent memory-pooling and chiplet-connectivity problems, but lack the rack-spanning, GPU-to-GPU bandwidth profile the prefill collectives studied here require \citep{Deloitte2026,cxl30spec,ucie20spec}. Silicon photonics and Co-Packaged Optics (CPO) reduce the cost of high-bandwidth datacenter links and have shown benefits for distributed MoE training \citep{lm2025moetraining}; this paper extends that analysis to inference, where quantization, continuous batching, disaggregated prefill/decode, and MLA-style KV compression shift the bottleneck among compute, memory, and collectives \citep{reuther2023ai,agrawal2024sarathi,zhong2024distserve,deepseekv3}.

\vspace{-8pt}
\section{Limitations}
\label{sec:limitations}

The results are analytical: we extend XLA's cost model with a projected optical layer not yet validated on deployed photonic hardware, and the optical cases assume ideal $4\times$ scale-up bandwidth without modeling additional link latency, thermal limits, signal-integrity effects, deployment cost, or full Total Cost of Ownership (TCO). The paper is also intentionally prefill-centric: the DES validation in Section~\ref{subsec:des_check} shows that decode saturation can dominate end-to-end serving behavior even when optical scale-up reduces TTFT. Consequently, the reported multipliers should be read as interconnect sensitivity results for prefill-heavy regimes: they quantify a real TTFT and input-throughput benefit from optical scale-up, but are not yet a complete serving-system sizing rule until decode capacity is co-designed alongside it.

\vspace{-8pt}
\section{Future Work}
\label{sec:future}

Future work should extend this prefill study into full serving-system co-design. Larger optical scale-up pods could support prefix-cache sharing, distributed KV-cache placement, faster prefill-to-decode KV transfer, and topology reconfiguration around failed GPUs; the key open question is how to jointly size optical prefill capacity and memory-bound decode capacity so TTFT reductions also translate into TPOT and end-to-end latency improvements. Running these workloads on optical hardware prototypes, rather than relying only on projected specifications, would help ground these projections.

\vspace{-8pt}
\section{Conclusion}
\label{sec:conclusion}

This paper presents modeled evidence that silicon-photonics-based optical interconnects can meaningfully benefit AI inference in communication-limited regimes. Across three MoE model scales and four hardware setups, the results suggest optical interconnects improve prefill latency and input-token throughput, particularly at high scale.

At 1K--8K tokens, the modeled results indicate 2.1--2.9$\times$ prefill improvements in high-batch configurations, where FP4 arithmetic shifts the critical path toward communication. At 128K tokens, rack-limited platforms (B200, B300, Rubin) see $\sim$4.3--5.8$\times$ gains by keeping 288+ GPUs inside a high-bandwidth scale-up pod rather than crossing the scale-out boundary. At 1M tokens, modeled improvements reach 2.2--4.5$\times$ on production platforms and up to 8.5$\times$ for the speculative R4 configuration. Optical scale-up matters most when it prevents a workload from crossing into a lower-bandwidth scale-out fabric, or when faster per-GPU arithmetic makes communication the limiting component.

\vspace{-4pt}
\section*{AI Assistance Disclosure}
AI tools assisted with editorial and formatting tasks. The authors reviewed and are responsible for all technical claims, data, citations, and conclusions in this submission.

\bibliography{references}

\appendices

\section{DeepSeek R1 Model Configuration}
\label{sec:model_impl}

We use a slightly modified version of the DeepSeek-R1 architecture to ensure divisibility constraints are met across all hardware mesh dimensions \citep{jaxllmexamples}. The embedding dimension is set to 8064, the number of attention heads to 144, the number of routed experts to 288, and the vocabulary size to 145,440. These values ensure divisibility by 8, 16, 24, 48, and 72 (the device counts used in the sweep). These modifications result in a model approximately 14\% larger than the production DeepSeek-R1 in terms of parameter count, but communication patterns and scaling behavior remain representative.

\begin{lstlisting}[language=Python,caption=DeepSeek R1 Model Configuration]
class Config:
  embed: int = 8064
  q_lora_rank: int = 1536
  kv_lora_rank: int = 512
  num_heads: int = 144
  qk_nope_head_dim: int = 128
  qk_rope_head_dim: int = 64
  v_head_dim: int = 128
  vocab_size: int = 145440
  num_layers: int = 61
  n_routed_experts: int = 288  # divisible by 72
  num_experts_per_tok: int = 8
  n_group: int = 8
  topk_group: int = 4
  moe_ffw_size: int = 2048
  ffw_size: int = 18432
  first_k_dense: int = 3
  n_shared_experts: int = 1
\end{lstlisting}

\section{Parallelism Strategy}
\label{sec:parallelism}

Our modeling supports 3D parallelism: Tensor (T), Expert (E), and Sequence (S) parallelism. The physical device mesh uses three axes:
\begin{itemize}
    \item Axis~\textbf{x}: splits the sequence dimension across racks (context parallelism).
    \item Axis~\textbf{y}: splits experts and attention heads within the scale-up pod (expert + tensor parallelism). This is the most bandwidth-intensive axis.
    \item Axis~\textbf{z}: splits the FFW hidden dimension of dense and expert layers (tensor parallelism).
\end{itemize}

For DeepSeek-R1 with 288 devices, the preferred mesh is $2\times144\times1$ (x$\times$y$\times$z): axis~y receives the largest allocation because both expert routing (all-to-all) and attention head partitioning (all-gather/reduce-scatter) benefit most from maximizing the y-dimension.

Table~\ref{tab:mesh_shapes} lists the selected mesh shape for each device count in the prefill sweep. Each configuration is the best-performing mesh found after evaluating hundreds of candidate shapes; the selection criterion is minimum overlapped prefill latency at the corresponding context length. Decode meshes are included for reference.

\begin{table}[htbp]
\centering
\footnotesize
\setlength{\tabcolsep}{5pt}
\caption{Selected prefill mesh shapes. Each entry is the best mesh out of hundreds of candidates evaluated for minimum overlapped prefill latency. Mesh dimensions are expressed as $x{\times}y{\times}z$, matching the \texttt{jax.make\_mesh} axis ordering \texttt{("x","y","z")} in the JAX LLM examples implementation \citep{jaxllmexamples}: $x$~=~sequence/context-parallel, $y$~=~expert+tensor-parallel (most bandwidth-intensive), $z$~=~FFW tensor-parallel.}
\label{tab:mesh_shapes}
\begin{tabularx}{\columnwidth}{@{}lXX@{}}
\toprule
\textbf{Phase} & \textbf{Devices} & \textbf{Mesh} ($x{\times}y{\times}z$) \\
\midrule
Prefill & 8    & $1\times8\times1$   \\
        & 16   & $1\times16\times1$  \\
        & 24   & $1\times24\times1$  \\
        & 48   & $1\times48\times1$  \\
        & 72   & $1\times72\times1$  \\
        & 144  & $2\times72\times1$  \\
        & 288  & $4\times72\times1$  \\
        & 576  & $8\times72\times1$  \\
        & 1152 & $16\times72\times1$ \\
\bottomrule
\end{tabularx}
\end{table}

\onecolumn
\section{GB200 Specifications}
\label{sec:gb200}

We derive hardware specifications from the OpenXLA project \citep{openxlagithub}. Memory capacity (192~GB HBM3e per GPU) and memory bandwidth (8~TB/s) are used unmodified. For compute, we use the \texttt{tcgen05.mma} tensor core instruction in 2-SM coordination mode, which supports $256\times256\times16$ shapes for BFloat16 (BF16) precision \citep{nv_blackwell_wp,nv_ptx_isa}. We apply precision-dependent multipliers: FP8 yields $2\times$ the BF16 rate; FP4 yields $4\times$ the BF16 rate.

\vspace{-8pt}
\section{Full Precision Sweep Heatmap (FP4 and FP8)}
\label{sec:full_heatmap}

Table~\ref{tab:sweep_heatmap_full} reproduces the sweep best multiplier heatmap from Section~\ref{subsec:summary} for both FP4 and FP8 quantization precisions. Within each model--hardware block, FP8 is listed first, followed by FP4. Column definitions are identical to Table~\ref{tab:sweep_heatmap}.

\begin{table}[H]
\centering
\caption{Full precision sweep best multiplier heatmap (FP8 and FP4). Column layout identical to Table~\ref{tab:sweep_heatmap}. Within each model--hardware block, FP8 is listed first and FP4 second. Cell color scales from min to max among these heatmap cells only.}
\label{tab:sweep_heatmap_full}
\scriptsize
\setlength{\tabcolsep}{2pt}
\renewcommand{\arraystretch}{1.28}
\resizebox{\textwidth}{!}{%
\begin{tabular}{llll|rrrrrrrrrrrrrrrrrrrrrrrrrrrrrrrrrrrr}
\toprule
Model & Hardware & Quant & & \multicolumn{36}{c}{Best multiplier (optical vs.\ no-optical)} \\
\cmidrule(lr){5-40}
 & & & seq\_len & \texttt{1K} & \texttt{1K} & \texttt{1K} & \texttt{1K} & \texttt{1K} & \texttt{1K} & \texttt{8K} & \texttt{8K} & \texttt{8K} & \texttt{8K} & \texttt{8K} & \texttt{8K} & \texttt{128K} & \texttt{128K} & \texttt{128K} & \texttt{128K} & \texttt{128K} & \texttt{128K} & \texttt{128K} & \texttt{128K} & \texttt{128K} & \texttt{1M} & \texttt{1M} & \texttt{1M} & \texttt{1M} & \texttt{1M} & \texttt{1M} & \texttt{1M} & \texttt{1M} & \texttt{1M} & \texttt{1M} & \texttt{1M} & \texttt{1M} & \texttt{1M} & \texttt{1M} & \texttt{1M} \\
 & & & tokens & \texttt{2M} & \texttt{8M} & \texttt{16M} & \texttt{2M} & \texttt{8M} & \texttt{16M} & \texttt{2M} & \texttt{8M} & \texttt{16M} & \texttt{2M} & \texttt{8M} & \texttt{16M} & \texttt{2M} & \texttt{8M} & \texttt{16M} & \texttt{2M} & \texttt{8M} & \texttt{16M} & \texttt{2M} & \texttt{8M} & \texttt{32M} & \texttt{2M} & \texttt{8M} & \texttt{16M} & \texttt{2M} & \texttt{8M} & \texttt{16M} & \texttt{2M} & \texttt{8M} & \texttt{32M} & \texttt{8M} & \texttt{32M} & \texttt{128M} & \texttt{8M} & \texttt{32M} & \texttt{128M} \\
 & & & dev & 16 & 16 & 16 & 72 & 72 & 72 & 16 & 16 & 16 & 72 & 72 & 72 & 16 & 16 & 16 & 72 & 72 & 72 & 288 & 288 & 288 & 16 & 16 & 16 & 72 & 72 & 72 & 288 & 288 & 288 & 576 & 576 & 576 & 1152 & 1152 & 1152 \\
 & & & batch & 2048 & 8192 & 16384 & 2048 & 8192 & 16384 & 256 & 1024 & 2048 & 256 & 1024 & 2048 & 16 & 64 & 128 & 16 & 64 & 128 & 16 & 64 & 256 & 2 & 8 & 16 & 2 & 8 & 16 & 2 & 8 & 32 & 8 & 32 & 64 & 8 & 32 & 64 \\
\midrule
mini & B200 & FP8 & & \cellcolor{heatblue!19}1.62 & \cellcolor{heatblue!21}1.84 & \cellcolor{heatblue!21}1.86 & \cellcolor{heatblue!20}1.80 & \cellcolor{heatblue!22}1.99 & \cellcolor{heatblue!23}2.01 & \cellcolor{heatblue!16}1.40 & \cellcolor{heatblue!17}1.49 & \cellcolor{heatblue!17}1.50 & \cellcolor{heatblue!19}1.67 & \cellcolor{heatblue!21}1.82 & \cellcolor{heatblue!21}1.83 & \cellcolor{heatblue!13}1.06 & \cellcolor{heatblue!13}1.06 & \cellcolor{heatblue!13}1.06 & \cellcolor{heatblue!14}1.20 & \cellcolor{heatblue!14}1.22 & \cellcolor{heatblue!14}1.22 & \cellcolor{heatblue!38}3.46 & \cellcolor{heatblue!44}3.98 & \cellcolor{heatblue!45}4.09 & \cellcolor{heatblue!12}1.01 & \cellcolor{heatblue!12}1.01 & \cellcolor{heatblue!12}1.01 & \cellcolor{heatblue!12}1.03 & \cellcolor{heatblue!12}1.03 & \cellcolor{heatblue!12}1.03 & \cellcolor{heatblue!17}1.47 & \cellcolor{heatblue!17}1.50 & \cellcolor{heatblue!17}1.50 & \cellcolor{heatblue!24}2.08 & \cellcolor{heatblue!24}2.11 & \cellcolor{heatblue!24}2.11 & \cellcolor{heatblue!36}3.20 & \cellcolor{heatblue!36}3.28 & \cellcolor{heatblue!36}3.29 \\
 &  & FP4 & & \cellcolor{heatblue!24}2.12 & \cellcolor{heatblue!25}2.23 & \cellcolor{heatblue!25}2.24 & \cellcolor{heatblue!26}2.33 & \cellcolor{heatblue!27}2.41 & \cellcolor{heatblue!27}2.41 & \cellcolor{heatblue!18}1.59 & \cellcolor{heatblue!19}1.61 & \cellcolor{heatblue!19}1.62 & \cellcolor{heatblue!23}2.04 & \cellcolor{heatblue!24}2.09 & \cellcolor{heatblue!24}2.09 & \cellcolor{heatblue!13}1.06 & \cellcolor{heatblue!13}1.07 & \cellcolor{heatblue!13}1.07 & \cellcolor{heatblue!14}1.23 & \cellcolor{heatblue!14}1.23 & \cellcolor{heatblue!14}1.23 & \cellcolor{heatblue!44}3.98 & \cellcolor{heatblue!47}4.27 & \cellcolor{heatblue!47}4.30 & \cellcolor{heatblue!12}1.01 & \cellcolor{heatblue!12}1.01 & \cellcolor{heatblue!12}1.01 & \cellcolor{heatblue!12}1.03 & \cellcolor{heatblue!12}1.03 & \cellcolor{heatblue!12}1.03 & \cellcolor{heatblue!17}1.50 & \cellcolor{heatblue!17}1.51 & \cellcolor{heatblue!17}1.51 & \cellcolor{heatblue!24}2.12 & \cellcolor{heatblue!24}2.13 & \cellcolor{heatblue!24}2.13 & \cellcolor{heatblue!36}3.29 & \cellcolor{heatblue!37}3.32 & \cellcolor{heatblue!37}3.32 \\
\addlinespace[2pt]
 & B300 & FP8 & & \cellcolor{heatblue!22}1.98 & \cellcolor{heatblue!24}2.12 & \cellcolor{heatblue!24}2.13 & \cellcolor{heatblue!24}2.16 & \cellcolor{heatblue!26}2.27 & \cellcolor{heatblue!26}2.28 & \cellcolor{heatblue!18}1.55 & \cellcolor{heatblue!18}1.59 & \cellcolor{heatblue!18}1.60 & \cellcolor{heatblue!22}1.93 & \cellcolor{heatblue!23}2.01 & \cellcolor{heatblue!23}2.02 & \cellcolor{heatblue!13}1.07 & \cellcolor{heatblue!13}1.07 & \cellcolor{heatblue!13}1.07 & \cellcolor{heatblue!14}1.23 & \cellcolor{heatblue!14}1.24 & \cellcolor{heatblue!14}1.24 & \cellcolor{heatblue!28}2.52 & \cellcolor{heatblue!31}2.74 & \cellcolor{heatblue!31}2.77 & \cellcolor{heatblue!12}1.01 & \cellcolor{heatblue!12}1.01 & \cellcolor{heatblue!12}1.01 & \cellcolor{heatblue!12}1.04 & \cellcolor{heatblue!12}1.04 & \cellcolor{heatblue!12}1.04 & \cellcolor{heatblue!15}1.27 & \cellcolor{heatblue!15}1.28 & \cellcolor{heatblue!15}1.28 & \cellcolor{heatblue!18}1.58 & \cellcolor{heatblue!18}1.59 & \cellcolor{heatblue!18}1.59 & \cellcolor{heatblue!24}2.17 & \cellcolor{heatblue!25}2.20 & \cellcolor{heatblue!25}2.20 \\
 &  & FP4 & & \cellcolor{heatblue!26}2.34 & \cellcolor{heatblue!27}2.41 & \cellcolor{heatblue!27}2.41 & \cellcolor{heatblue!29}2.59 & \cellcolor{heatblue!29}2.64 & \cellcolor{heatblue!30}2.64 & \cellcolor{heatblue!19}1.65 & \cellcolor{heatblue!19}1.67 & \cellcolor{heatblue!19}1.67 & \cellcolor{heatblue!25}2.21 & \cellcolor{heatblue!25}2.24 & \cellcolor{heatblue!25}2.24 & \cellcolor{heatblue!13}1.07 & \cellcolor{heatblue!13}1.07 & \cellcolor{heatblue!13}1.07 & \cellcolor{heatblue!15}1.25 & \cellcolor{heatblue!15}1.25 & \cellcolor{heatblue!15}1.25 & \cellcolor{heatblue!31}2.75 & \cellcolor{heatblue!32}2.84 & \cellcolor{heatblue!32}2.85 & \cellcolor{heatblue!12}1.01 & \cellcolor{heatblue!12}1.01 & \cellcolor{heatblue!12}1.01 & \cellcolor{heatblue!12}1.04 & \cellcolor{heatblue!12}1.04 & \cellcolor{heatblue!12}1.04 & \cellcolor{heatblue!15}1.28 & \cellcolor{heatblue!15}1.28 & \cellcolor{heatblue!15}1.28 & \cellcolor{heatblue!18}1.60 & \cellcolor{heatblue!18}1.60 & \cellcolor{heatblue!18}1.60 & \cellcolor{heatblue!25}2.20 & \cellcolor{heatblue!25}2.21 & \cellcolor{heatblue!25}2.21 \\
\addlinespace[2pt]
 & Rubin & FP8 & & \cellcolor{heatblue!25}2.24 & \cellcolor{heatblue!27}2.41 & \cellcolor{heatblue!27}2.43 & \cellcolor{heatblue!27}2.43 & \cellcolor{heatblue!29}2.56 & \cellcolor{heatblue!29}2.57 & \cellcolor{heatblue!20}1.73 & \cellcolor{heatblue!20}1.78 & \cellcolor{heatblue!20}1.79 & \cellcolor{heatblue!25}2.19 & \cellcolor{heatblue!26}2.28 & \cellcolor{heatblue!26}2.29 & \cellcolor{heatblue!13}1.09 & \cellcolor{heatblue!13}1.09 & \cellcolor{heatblue!13}1.09 & \cellcolor{heatblue!15}1.32 & \cellcolor{heatblue!15}1.32 & \cellcolor{heatblue!15}1.33 & \cellcolor{heatblue!53}4.82 & \cellcolor{heatblue!59}5.35 & \cellcolor{heatblue!59}5.42 & \cellcolor{heatblue!12}1.01 & \cellcolor{heatblue!12}1.01 & \cellcolor{heatblue!12}1.01 & \cellcolor{heatblue!12}1.05 & \cellcolor{heatblue!12}1.05 & \cellcolor{heatblue!12}1.05 & \cellcolor{heatblue!19}1.70 & \cellcolor{heatblue!20}1.72 & \cellcolor{heatblue!20}1.72 & \cellcolor{heatblue!29}2.57 & \cellcolor{heatblue!29}2.59 & \cellcolor{heatblue!29}2.59 & \cellcolor{heatblue!46}4.18 & \cellcolor{heatblue!47}4.25 & \cellcolor{heatblue!47}4.26 \\
 &  & FP4 & & \cellcolor{heatblue!28}2.48 & \cellcolor{heatblue!29}2.62 & \cellcolor{heatblue!29}2.62 & \cellcolor{heatblue!30}2.71 & \cellcolor{heatblue!31}2.80 & \cellcolor{heatblue!31}2.81 & \cellcolor{heatblue!21}1.81 & \cellcolor{heatblue!21}1.85 & \cellcolor{heatblue!21}1.85 & \cellcolor{heatblue!27}2.38 & \cellcolor{heatblue!27}2.45 & \cellcolor{heatblue!27}2.45 & \cellcolor{heatblue!13}1.10 & \cellcolor{heatblue!13}1.10 & \cellcolor{heatblue!13}1.10 & \cellcolor{heatblue!15}1.33 & \cellcolor{heatblue!16}1.34 & \cellcolor{heatblue!16}1.34 & \cellcolor{heatblue!57}5.16 & \cellcolor{heatblue!60}5.51 & \cellcolor{heatblue!61}5.55 & \cellcolor{heatblue!12}1.01 & \cellcolor{heatblue!12}1.01 & \cellcolor{heatblue!12}1.01 & \cellcolor{heatblue!12}1.05 & \cellcolor{heatblue!12}1.05 & \cellcolor{heatblue!12}1.05 & \cellcolor{heatblue!20}1.71 & \cellcolor{heatblue!20}1.72 & \cellcolor{heatblue!20}1.73 & \cellcolor{heatblue!29}2.59 & \cellcolor{heatblue!29}2.60 & \cellcolor{heatblue!29}2.60 & \cellcolor{heatblue!47}4.24 & \cellcolor{heatblue!47}4.28 & \cellcolor{heatblue!47}4.28 \\
\addlinespace[2pt]
 & R4 & FP8 & & \cellcolor{heatblue!31}2.79 & \cellcolor{heatblue!33}2.94 & \cellcolor{heatblue!33}2.95 & \cellcolor{heatblue!32}2.90 & \cellcolor{heatblue!33}3.00 & \cellcolor{heatblue!34}3.01 & \cellcolor{heatblue!25}2.25 & \cellcolor{heatblue!26}2.33 & \cellcolor{heatblue!26}2.34 & \cellcolor{heatblue!31}2.74 & \cellcolor{heatblue!32}2.83 & \cellcolor{heatblue!32}2.84 & \cellcolor{heatblue!14}1.21 & \cellcolor{heatblue!14}1.21 & \cellcolor{heatblue!14}1.21 & \cellcolor{heatblue!19}1.64 & \cellcolor{heatblue!19}1.65 & \cellcolor{heatblue!19}1.65 & \cellcolor{heatblue!21}1.86 & \cellcolor{heatblue!22}1.95 & \cellcolor{heatblue!22}1.96 & \cellcolor{heatblue!12}1.03 & \cellcolor{heatblue!12}1.03 & \cellcolor{heatblue!12}1.03 & \cellcolor{heatblue!13}1.11 & \cellcolor{heatblue!13}1.11 & \cellcolor{heatblue!13}1.11 & \cellcolor{heatblue!14}1.19 & \cellcolor{heatblue!14}1.19 & \cellcolor{heatblue!14}1.19 & \cellcolor{heatblue!15}1.28 & \cellcolor{heatblue!15}1.28 & \cellcolor{heatblue!15}1.28 & \cellcolor{heatblue!86}7.95 & \cellcolor{heatblue!88}8.08 & \cellcolor{heatblue!88}8.10 \\
 &  & FP4 & & \cellcolor{heatblue!33}3.00 & \cellcolor{heatblue!34}3.10 & \cellcolor{heatblue!35}3.11 & \cellcolor{heatblue!35}3.11 & \cellcolor{heatblue!35}3.18 & \cellcolor{heatblue!35}3.19 & \cellcolor{heatblue!26}2.36 & \cellcolor{heatblue!27}2.41 & \cellcolor{heatblue!27}2.42 & \cellcolor{heatblue!33}2.92 & \cellcolor{heatblue!33}2.98 & \cellcolor{heatblue!33}2.99 & \cellcolor{heatblue!14}1.21 & \cellcolor{heatblue!14}1.21 & \cellcolor{heatblue!14}1.22 & \cellcolor{heatblue!19}1.66 & \cellcolor{heatblue!19}1.67 & \cellcolor{heatblue!19}1.67 & \cellcolor{heatblue!22}1.91 & \cellcolor{heatblue!22}1.98 & \cellcolor{heatblue!23}1.99 & \cellcolor{heatblue!12}1.03 & \cellcolor{heatblue!12}1.03 & \cellcolor{heatblue!12}1.03 & \cellcolor{heatblue!13}1.11 & \cellcolor{heatblue!13}1.11 & \cellcolor{heatblue!13}1.11 & \cellcolor{heatblue!14}1.19 & \cellcolor{heatblue!14}1.19 & \cellcolor{heatblue!14}1.19 & \cellcolor{heatblue!15}1.28 & \cellcolor{heatblue!15}1.28 & \cellcolor{heatblue!15}1.28 & \cellcolor{heatblue!87}8.04 & \cellcolor{heatblue!89}8.14 & \cellcolor{heatblue!89}8.15 \\
\midrule
r1 & B200 & FP8 & & \cellcolor{heatblue!19}1.63 & \cellcolor{heatblue!21}1.84 & \cellcolor{heatblue!21}1.85 & \cellcolor{heatblue!20}1.78 & \cellcolor{heatblue!22}1.94 & \cellcolor{heatblue!22}1.97 & \cellcolor{heatblue!16}1.41 & \cellcolor{heatblue!17}1.48 & \cellcolor{heatblue!17}1.50 & \cellcolor{heatblue!19}1.67 & \cellcolor{heatblue!20}1.77 & \cellcolor{heatblue!21}1.81 & \cellcolor{heatblue!13}1.06 & \cellcolor{heatblue!13}1.06 & \cellcolor{heatblue!13}1.06 & \cellcolor{heatblue!14}1.20 & \cellcolor{heatblue!14}1.22 & \cellcolor{heatblue!14}1.22 & \cellcolor{heatblue!39}3.57 & \cellcolor{heatblue!45}4.11 & \cellcolor{heatblue!46}4.22 & \cellcolor{heatblue!12}1.01 & \cellcolor{heatblue!12}1.01 & \cellcolor{heatblue!12}1.01 & \cellcolor{heatblue!12}1.03 & \cellcolor{heatblue!12}1.03 & \cellcolor{heatblue!12}1.03 & \cellcolor{heatblue!17}1.50 & \cellcolor{heatblue!18}1.52 & \cellcolor{heatblue!18}1.53 & \cellcolor{heatblue!24}2.14 & \cellcolor{heatblue!24}2.17 & \cellcolor{heatblue!24}2.17 & \cellcolor{heatblue!37}3.32 & \cellcolor{heatblue!38}3.40 & \cellcolor{heatblue!38}3.41 \\
 &  & FP4 & & \cellcolor{heatblue!24}2.11 & \cellcolor{heatblue!25}2.21 & \cellcolor{heatblue!25}2.22 & \cellcolor{heatblue!26}2.28 & \cellcolor{heatblue!26}2.35 & \cellcolor{heatblue!26}2.36 & \cellcolor{heatblue!18}1.59 & \cellcolor{heatblue!19}1.62 & \cellcolor{heatblue!19}1.62 & \cellcolor{heatblue!23}2.01 & \cellcolor{heatblue!23}2.06 & \cellcolor{heatblue!23}2.06 & \cellcolor{heatblue!13}1.07 & \cellcolor{heatblue!13}1.07 & \cellcolor{heatblue!13}1.07 & \cellcolor{heatblue!14}1.23 & \cellcolor{heatblue!14}1.23 & \cellcolor{heatblue!14}1.23 & \cellcolor{heatblue!45}4.12 & \cellcolor{heatblue!49}4.42 & \cellcolor{heatblue!49}4.46 & \cellcolor{heatblue!12}1.01 & \cellcolor{heatblue!12}1.01 & \cellcolor{heatblue!12}1.01 & \cellcolor{heatblue!12}1.03 & \cellcolor{heatblue!12}1.03 & \cellcolor{heatblue!12}1.03 & \cellcolor{heatblue!18}1.53 & \cellcolor{heatblue!18}1.53 & \cellcolor{heatblue!18}1.54 & \cellcolor{heatblue!25}2.18 & \cellcolor{heatblue!25}2.19 & \cellcolor{heatblue!25}2.19 & \cellcolor{heatblue!38}3.41 & \cellcolor{heatblue!38}3.45 & \cellcolor{heatblue!38}3.45 \\
\addlinespace[2pt]
 & B300 & FP8 & & \cellcolor{heatblue!22}1.98 & \cellcolor{heatblue!24}2.10 & \cellcolor{heatblue!24}2.12 & \cellcolor{heatblue!24}2.12 & \cellcolor{heatblue!25}2.21 & \cellcolor{heatblue!25}2.22 & \cellcolor{heatblue!18}1.56 & \cellcolor{heatblue!18}1.58 & \cellcolor{heatblue!18}1.60 & \cellcolor{heatblue!22}1.91 & \cellcolor{heatblue!22}1.95 & \cellcolor{heatblue!23}1.99 & \cellcolor{heatblue!13}1.07 & \cellcolor{heatblue!13}1.07 & \cellcolor{heatblue!13}1.07 & \cellcolor{heatblue!14}1.23 & \cellcolor{heatblue!14}1.23 & \cellcolor{heatblue!14}1.23 & \cellcolor{heatblue!29}2.59 & \cellcolor{heatblue!31}2.82 & \cellcolor{heatblue!32}2.85 & \cellcolor{heatblue!12}1.01 & \cellcolor{heatblue!12}1.01 & \cellcolor{heatblue!12}1.01 & \cellcolor{heatblue!12}1.03 & \cellcolor{heatblue!12}1.04 & \cellcolor{heatblue!12}1.04 & \cellcolor{heatblue!15}1.28 & \cellcolor{heatblue!15}1.29 & \cellcolor{heatblue!15}1.29 & \cellcolor{heatblue!19}1.62 & \cellcolor{heatblue!19}1.63 & \cellcolor{heatblue!19}1.63 & \cellcolor{heatblue!25}2.24 & \cellcolor{heatblue!26}2.27 & \cellcolor{heatblue!26}2.27 \\
 &  & FP4 & & \cellcolor{heatblue!26}2.32 & \cellcolor{heatblue!27}2.38 & \cellcolor{heatblue!27}2.38 & \cellcolor{heatblue!28}2.53 & \cellcolor{heatblue!29}2.58 & \cellcolor{heatblue!29}2.58 & \cellcolor{heatblue!19}1.65 & \cellcolor{heatblue!19}1.67 & \cellcolor{heatblue!19}1.67 & \cellcolor{heatblue!25}2.18 & \cellcolor{heatblue!25}2.21 & \cellcolor{heatblue!25}2.21 & \cellcolor{heatblue!13}1.07 & \cellcolor{heatblue!13}1.07 & \cellcolor{heatblue!13}1.07 & \cellcolor{heatblue!15}1.24 & \cellcolor{heatblue!15}1.25 & \cellcolor{heatblue!15}1.25 & \cellcolor{heatblue!32}2.83 & \cellcolor{heatblue!33}2.93 & \cellcolor{heatblue!33}2.93 & \cellcolor{heatblue!12}1.01 & \cellcolor{heatblue!12}1.01 & \cellcolor{heatblue!12}1.01 & \cellcolor{heatblue!12}1.04 & \cellcolor{heatblue!12}1.04 & \cellcolor{heatblue!12}1.04 & \cellcolor{heatblue!15}1.29 & \cellcolor{heatblue!15}1.30 & \cellcolor{heatblue!15}1.30 & \cellcolor{heatblue!19}1.63 & \cellcolor{heatblue!19}1.63 & \cellcolor{heatblue!19}1.63 & \cellcolor{heatblue!26}2.27 & \cellcolor{heatblue!26}2.28 & \cellcolor{heatblue!26}2.28 \\
\addlinespace[2pt]
 & Rubin & FP8 & & \cellcolor{heatblue!25}2.23 & \cellcolor{heatblue!27}2.38 & \cellcolor{heatblue!27}2.40 & \cellcolor{heatblue!27}2.37 & \cellcolor{heatblue!28}2.49 & \cellcolor{heatblue!28}2.50 & \cellcolor{heatblue!20}1.74 & \cellcolor{heatblue!20}1.76 & \cellcolor{heatblue!20}1.79 & \cellcolor{heatblue!24}2.16 & \cellcolor{heatblue!25}2.21 & \cellcolor{heatblue!25}2.25 & \cellcolor{heatblue!13}1.09 & \cellcolor{heatblue!13}1.10 & \cellcolor{heatblue!13}1.10 & \cellcolor{heatblue!15}1.32 & \cellcolor{heatblue!15}1.32 & \cellcolor{heatblue!15}1.32 & \cellcolor{heatblue!55}4.99 & \cellcolor{heatblue!61}5.54 & \cellcolor{heatblue!61}5.62 & \cellcolor{heatblue!12}1.01 & \cellcolor{heatblue!12}1.01 & \cellcolor{heatblue!12}1.01 & \cellcolor{heatblue!12}1.05 & \cellcolor{heatblue!12}1.05 & \cellcolor{heatblue!12}1.05 & \cellcolor{heatblue!20}1.74 & \cellcolor{heatblue!20}1.76 & \cellcolor{heatblue!20}1.76 & \cellcolor{heatblue!30}2.66 & \cellcolor{heatblue!30}2.68 & \cellcolor{heatblue!30}2.68 & \cellcolor{heatblue!48}4.35 & \cellcolor{heatblue!49}4.42 & \cellcolor{heatblue!49}4.43 \\
 &  & FP4 & & \cellcolor{heatblue!28}2.45 & \cellcolor{heatblue!29}2.58 & \cellcolor{heatblue!29}2.58 & \cellcolor{heatblue!30}2.64 & \cellcolor{heatblue!30}2.72 & \cellcolor{heatblue!30}2.73 & \cellcolor{heatblue!21}1.81 & \cellcolor{heatblue!21}1.85 & \cellcolor{heatblue!21}1.85 & \cellcolor{heatblue!26}2.34 & \cellcolor{heatblue!27}2.40 & \cellcolor{heatblue!27}2.40 & \cellcolor{heatblue!13}1.10 & \cellcolor{heatblue!13}1.10 & \cellcolor{heatblue!13}1.10 & \cellcolor{heatblue!15}1.33 & \cellcolor{heatblue!15}1.33 & \cellcolor{heatblue!15}1.33 & \cellcolor{heatblue!58}5.34 & \cellcolor{heatblue!62}5.71 & \cellcolor{heatblue!63}5.75 & \cellcolor{heatblue!12}1.01 & \cellcolor{heatblue!12}1.01 & \cellcolor{heatblue!12}1.01 & \cellcolor{heatblue!12}1.05 & \cellcolor{heatblue!12}1.05 & \cellcolor{heatblue!12}1.05 & \cellcolor{heatblue!20}1.75 & \cellcolor{heatblue!20}1.76 & \cellcolor{heatblue!20}1.76 & \cellcolor{heatblue!30}2.68 & \cellcolor{heatblue!30}2.69 & \cellcolor{heatblue!30}2.69 & \cellcolor{heatblue!48}4.41 & \cellcolor{heatblue!49}4.45 & \cellcolor{heatblue!49}4.45 \\
\addlinespace[2pt]
 & R4 & FP8 & & \cellcolor{heatblue!31}2.74 & \cellcolor{heatblue!32}2.88 & \cellcolor{heatblue!32}2.90 & \cellcolor{heatblue!32}2.83 & \cellcolor{heatblue!33}2.94 & \cellcolor{heatblue!33}2.95 & \cellcolor{heatblue!25}2.24 & \cellcolor{heatblue!26}2.28 & \cellcolor{heatblue!26}2.32 & \cellcolor{heatblue!30}2.68 & \cellcolor{heatblue!30}2.73 & \cellcolor{heatblue!31}2.77 & \cellcolor{heatblue!14}1.21 & \cellcolor{heatblue!14}1.21 & \cellcolor{heatblue!14}1.21 & \cellcolor{heatblue!19}1.63 & \cellcolor{heatblue!19}1.64 & \cellcolor{heatblue!19}1.65 & \cellcolor{heatblue!21}1.88 & \cellcolor{heatblue!22}1.98 & \cellcolor{heatblue!23}2.00 & \cellcolor{heatblue!12}1.03 & \cellcolor{heatblue!12}1.03 & \cellcolor{heatblue!12}1.03 & \cellcolor{heatblue!13}1.11 & \cellcolor{heatblue!13}1.11 & \cellcolor{heatblue!13}1.11 & \cellcolor{heatblue!14}1.20 & \cellcolor{heatblue!14}1.20 & \cellcolor{heatblue!14}1.20 & \cellcolor{heatblue!15}1.30 & \cellcolor{heatblue!15}1.30 & \cellcolor{heatblue!15}1.31 & \cellcolor{heatblue!89}8.22 & \cellcolor{heatblue!91}8.37 & \cellcolor{heatblue!91}8.38 \\
 &  & FP4 & & \cellcolor{heatblue!33}2.95 & \cellcolor{heatblue!34}3.05 & \cellcolor{heatblue!34}3.06 & \cellcolor{heatblue!34}3.05 & \cellcolor{heatblue!35}3.13 & \cellcolor{heatblue!35}3.13 & \cellcolor{heatblue!26}2.34 & \cellcolor{heatblue!27}2.39 & \cellcolor{heatblue!27}2.40 & \cellcolor{heatblue!32}2.86 & \cellcolor{heatblue!32}2.92 & \cellcolor{heatblue!33}2.92 & \cellcolor{heatblue!14}1.21 & \cellcolor{heatblue!14}1.22 & \cellcolor{heatblue!14}1.22 & \cellcolor{heatblue!19}1.66 & \cellcolor{heatblue!19}1.67 & \cellcolor{heatblue!19}1.67 & \cellcolor{heatblue!22}1.94 & \cellcolor{heatblue!23}2.01 & \cellcolor{heatblue!23}2.02 & \cellcolor{heatblue!12}1.03 & \cellcolor{heatblue!12}1.03 & \cellcolor{heatblue!12}1.03 & \cellcolor{heatblue!13}1.11 & \cellcolor{heatblue!13}1.11 & \cellcolor{heatblue!13}1.11 & \cellcolor{heatblue!14}1.20 & \cellcolor{heatblue!14}1.20 & \cellcolor{heatblue!14}1.20 & \cellcolor{heatblue!15}1.31 & \cellcolor{heatblue!15}1.31 & \cellcolor{heatblue!15}1.31 & \cellcolor{heatblue!91}8.33 & \cellcolor{heatblue!92}8.43 & \cellcolor{heatblue!92}8.44 \\
\midrule
next & B200 & FP8 & & \cellcolor{heatblue!20}1.73 & \cellcolor{heatblue!20}1.79 & \cellcolor{heatblue!20}1.80 & \cellcolor{heatblue!22}1.91 & \cellcolor{heatblue!22}1.96 & \cellcolor{heatblue!22}1.97 & \cellcolor{heatblue!17}1.45 & \cellcolor{heatblue!17}1.47 & \cellcolor{heatblue!17}1.48 & \cellcolor{heatblue!20}1.78 & \cellcolor{heatblue!21}1.81 & \cellcolor{heatblue!21}1.82 & \cellcolor{heatblue!13}1.06 & \cellcolor{heatblue!13}1.06 & \cellcolor{heatblue!13}1.06 & \cellcolor{heatblue!14}1.22 & \cellcolor{heatblue!14}1.22 & \cellcolor{heatblue!14}1.22 & \cellcolor{heatblue!43}3.91 & \cellcolor{heatblue!47}4.27 & \cellcolor{heatblue!47}4.30 & \cellcolor{heatblue!12}1.01 & \cellcolor{heatblue!12}1.01 & \cellcolor{heatblue!12}1.01 & \cellcolor{heatblue!12}1.03 & \cellcolor{heatblue!12}1.03 & \cellcolor{heatblue!12}1.03 & \cellcolor{heatblue!18}1.53 & \cellcolor{heatblue!18}1.54 & \cellcolor{heatblue!18}1.54 & \cellcolor{heatblue!25}2.20 & \cellcolor{heatblue!25}2.21 & \cellcolor{heatblue!25}2.21 & \cellcolor{heatblue!38}3.42 & \cellcolor{heatblue!38}3.47 & \cellcolor{heatblue!38}3.47 \\
 &  & FP4 & & \cellcolor{heatblue!25}2.18 & \cellcolor{heatblue!25}2.23 & \cellcolor{heatblue!25}2.23 & \cellcolor{heatblue!26}2.33 & \cellcolor{heatblue!26}2.36 & \cellcolor{heatblue!27}2.36 & \cellcolor{heatblue!19}1.62 & \cellcolor{heatblue!19}1.63 & \cellcolor{heatblue!19}1.63 & \cellcolor{heatblue!23}2.05 & \cellcolor{heatblue!23}2.07 & \cellcolor{heatblue!23}2.08 & \cellcolor{heatblue!13}1.07 & \cellcolor{heatblue!13}1.07 & \cellcolor{heatblue!13}1.07 & \cellcolor{heatblue!14}1.24 & \cellcolor{heatblue!14}1.24 & \cellcolor{heatblue!14}1.24 & \cellcolor{heatblue!48}4.38 & \cellcolor{heatblue!50}4.52 & \cellcolor{heatblue!50}4.53 & \cellcolor{heatblue!12}1.01 & \cellcolor{heatblue!12}1.01 & \cellcolor{heatblue!12}1.01 & \cellcolor{heatblue!12}1.03 & \cellcolor{heatblue!12}1.03 & \cellcolor{heatblue!12}1.03 & \cellcolor{heatblue!18}1.55 & \cellcolor{heatblue!18}1.55 & \cellcolor{heatblue!18}1.55 & \cellcolor{heatblue!25}2.22 & \cellcolor{heatblue!25}2.23 & \cellcolor{heatblue!25}2.23 & \cellcolor{heatblue!39}3.50 & \cellcolor{heatblue!39}3.51 & \cellcolor{heatblue!39}3.51 \\
\addlinespace[2pt]
 & B300 & FP8 & & \cellcolor{heatblue!23}2.08 & \cellcolor{heatblue!24}2.12 & \cellcolor{heatblue!24}2.13 & \cellcolor{heatblue!25}2.19 & \cellcolor{heatblue!25}2.22 & \cellcolor{heatblue!25}2.23 & \cellcolor{heatblue!18}1.59 & \cellcolor{heatblue!18}1.61 & \cellcolor{heatblue!18}1.61 & \cellcolor{heatblue!22}1.97 & \cellcolor{heatblue!23}1.99 & \cellcolor{heatblue!23}2.00 & \cellcolor{heatblue!13}1.07 & \cellcolor{heatblue!13}1.07 & \cellcolor{heatblue!13}1.07 & \cellcolor{heatblue!14}1.24 & \cellcolor{heatblue!14}1.24 & \cellcolor{heatblue!14}1.24 & \cellcolor{heatblue!31}2.77 & \cellcolor{heatblue!32}2.89 & \cellcolor{heatblue!32}2.90 & \cellcolor{heatblue!12}1.01 & \cellcolor{heatblue!12}1.01 & \cellcolor{heatblue!12}1.01 & \cellcolor{heatblue!12}1.04 & \cellcolor{heatblue!12}1.04 & \cellcolor{heatblue!12}1.04 & \cellcolor{heatblue!15}1.30 & \cellcolor{heatblue!15}1.30 & \cellcolor{heatblue!15}1.30 & \cellcolor{heatblue!19}1.65 & \cellcolor{heatblue!19}1.65 & \cellcolor{heatblue!19}1.65 & \cellcolor{heatblue!26}2.30 & \cellcolor{heatblue!26}2.32 & \cellcolor{heatblue!26}2.32 \\
 &  & FP4 & & \cellcolor{heatblue!27}2.38 & \cellcolor{heatblue!27}2.40 & \cellcolor{heatblue!27}2.40 & \cellcolor{heatblue!29}2.57 & \cellcolor{heatblue!29}2.59 & \cellcolor{heatblue!29}2.59 & \cellcolor{heatblue!19}1.68 & \cellcolor{heatblue!19}1.68 & \cellcolor{heatblue!19}1.68 & \cellcolor{heatblue!25}2.21 & \cellcolor{heatblue!25}2.23 & \cellcolor{heatblue!25}2.23 & \cellcolor{heatblue!13}1.07 & \cellcolor{heatblue!13}1.07 & \cellcolor{heatblue!13}1.07 & \cellcolor{heatblue!15}1.25 & \cellcolor{heatblue!15}1.25 & \cellcolor{heatblue!15}1.25 & \cellcolor{heatblue!33}2.95 & \cellcolor{heatblue!33}2.99 & \cellcolor{heatblue!33}2.99 & \cellcolor{heatblue!12}1.01 & \cellcolor{heatblue!12}1.01 & \cellcolor{heatblue!12}1.01 & \cellcolor{heatblue!12}1.04 & \cellcolor{heatblue!12}1.04 & \cellcolor{heatblue!12}1.04 & \cellcolor{heatblue!15}1.31 & \cellcolor{heatblue!15}1.31 & \cellcolor{heatblue!15}1.31 & \cellcolor{heatblue!19}1.66 & \cellcolor{heatblue!19}1.66 & \cellcolor{heatblue!19}1.66 & \cellcolor{heatblue!26}2.33 & \cellcolor{heatblue!26}2.33 & \cellcolor{heatblue!26}2.33 \\
\addlinespace[2pt]
 & Rubin & FP8 & & \cellcolor{heatblue!26}2.34 & \cellcolor{heatblue!27}2.40 & \cellcolor{heatblue!27}2.40 & \cellcolor{heatblue!27}2.45 & \cellcolor{heatblue!28}2.50 & \cellcolor{heatblue!28}2.51 & \cellcolor{heatblue!20}1.78 & \cellcolor{heatblue!21}1.80 & \cellcolor{heatblue!21}1.80 & \cellcolor{heatblue!25}2.22 & \cellcolor{heatblue!25}2.25 & \cellcolor{heatblue!25}2.26 & \cellcolor{heatblue!13}1.10 & \cellcolor{heatblue!13}1.10 & \cellcolor{heatblue!13}1.10 & \cellcolor{heatblue!15}1.33 & \cellcolor{heatblue!15}1.33 & \cellcolor{heatblue!15}1.33 & \cellcolor{heatblue!59}5.40 & \cellcolor{heatblue!62}5.67 & \cellcolor{heatblue!62}5.70 & \cellcolor{heatblue!12}1.01 & \cellcolor{heatblue!12}1.01 & \cellcolor{heatblue!12}1.01 & \cellcolor{heatblue!12}1.05 & \cellcolor{heatblue!12}1.05 & \cellcolor{heatblue!12}1.05 & \cellcolor{heatblue!20}1.77 & \cellcolor{heatblue!20}1.78 & \cellcolor{heatblue!20}1.78 & \cellcolor{heatblue!30}2.72 & \cellcolor{heatblue!30}2.73 & \cellcolor{heatblue!30}2.73 & \cellcolor{heatblue!49}4.47 & \cellcolor{heatblue!50}4.51 & \cellcolor{heatblue!50}4.51 \\
 &  & FP4 & & \cellcolor{heatblue!28}2.53 & \cellcolor{heatblue!29}2.58 & \cellcolor{heatblue!29}2.59 & \cellcolor{heatblue!30}2.69 & \cellcolor{heatblue!30}2.73 & \cellcolor{heatblue!31}2.73 & \cellcolor{heatblue!21}1.85 & \cellcolor{heatblue!21}1.86 & \cellcolor{heatblue!21}1.86 & \cellcolor{heatblue!27}2.38 & \cellcolor{heatblue!27}2.41 & \cellcolor{heatblue!27}2.41 & \cellcolor{heatblue!13}1.10 & \cellcolor{heatblue!13}1.10 & \cellcolor{heatblue!13}1.10 & \cellcolor{heatblue!16}1.34 & \cellcolor{heatblue!16}1.34 & \cellcolor{heatblue!16}1.34 & \cellcolor{heatblue!62}5.67 & \cellcolor{heatblue!64}5.82 & \cellcolor{heatblue!64}5.84 & \cellcolor{heatblue!12}1.01 & \cellcolor{heatblue!12}1.01 & \cellcolor{heatblue!12}1.01 & \cellcolor{heatblue!12}1.05 & \cellcolor{heatblue!12}1.05 & \cellcolor{heatblue!12}1.05 & \cellcolor{heatblue!20}1.79 & \cellcolor{heatblue!20}1.79 & \cellcolor{heatblue!20}1.79 & \cellcolor{heatblue!31}2.74 & \cellcolor{heatblue!31}2.74 & \cellcolor{heatblue!31}2.74 & \cellcolor{heatblue!50}4.51 & \cellcolor{heatblue!50}4.53 & \cellcolor{heatblue!50}4.53 \\
\addlinespace[2pt]
 & R4 & FP8 & & \cellcolor{heatblue!32}2.83 & \cellcolor{heatblue!32}2.89 & \cellcolor{heatblue!32}2.89 & \cellcolor{heatblue!32}2.89 & \cellcolor{heatblue!33}2.94 & \cellcolor{heatblue!33}2.95 & \cellcolor{heatblue!26}2.29 & \cellcolor{heatblue!26}2.32 & \cellcolor{heatblue!26}2.33 & \cellcolor{heatblue!30}2.73 & \cellcolor{heatblue!31}2.76 & \cellcolor{heatblue!31}2.77 & \cellcolor{heatblue!14}1.22 & \cellcolor{heatblue!14}1.22 & \cellcolor{heatblue!14}1.22 & \cellcolor{heatblue!19}1.65 & \cellcolor{heatblue!19}1.66 & \cellcolor{heatblue!19}1.66 & \cellcolor{heatblue!23}2.02 & \cellcolor{heatblue!23}2.06 & \cellcolor{heatblue!23}2.07 & \cellcolor{heatblue!12}1.03 & \cellcolor{heatblue!12}1.03 & \cellcolor{heatblue!12}1.03 & \cellcolor{heatblue!13}1.12 & \cellcolor{heatblue!13}1.12 & \cellcolor{heatblue!13}1.12 & \cellcolor{heatblue!14}1.22 & \cellcolor{heatblue!14}1.22 & \cellcolor{heatblue!14}1.22 & \cellcolor{heatblue!16}1.35 & \cellcolor{heatblue!16}1.35 & \cellcolor{heatblue!16}1.35 & \cellcolor{heatblue!91}8.35 & \cellcolor{heatblue!91}8.41 & \cellcolor{heatblue!91}8.41 \\
 &  & FP4 & & \cellcolor{heatblue!33}3.01 & \cellcolor{heatblue!34}3.05 & \cellcolor{heatblue!34}3.05 & \cellcolor{heatblue!34}3.10 & \cellcolor{heatblue!35}3.13 & \cellcolor{heatblue!35}3.13 & \cellcolor{heatblue!27}2.38 & \cellcolor{heatblue!27}2.40 & \cellcolor{heatblue!27}2.40 & \cellcolor{heatblue!32}2.89 & \cellcolor{heatblue!32}2.91 & \cellcolor{heatblue!32}2.92 & \cellcolor{heatblue!14}1.22 & \cellcolor{heatblue!14}1.22 & \cellcolor{heatblue!14}1.22 & \cellcolor{heatblue!19}1.67 & \cellcolor{heatblue!19}1.68 & \cellcolor{heatblue!19}1.68 & \cellcolor{heatblue!23}2.06 & \cellcolor{heatblue!24}2.09 & \cellcolor{heatblue!24}2.10 & \cellcolor{heatblue!12}1.03 & \cellcolor{heatblue!12}1.03 & \cellcolor{heatblue!12}1.03 & \cellcolor{heatblue!13}1.12 & \cellcolor{heatblue!13}1.12 & \cellcolor{heatblue!13}1.12 & \cellcolor{heatblue!14}1.22 & \cellcolor{heatblue!14}1.22 & \cellcolor{heatblue!14}1.22 & \cellcolor{heatblue!16}1.35 & \cellcolor{heatblue!16}1.35 & \cellcolor{heatblue!16}1.35 & \cellcolor{heatblue!92}8.42 & \cellcolor{heatblue!92}8.46 & \cellcolor{heatblue!92}8.47 \\
\bottomrule
\end{tabular}
}
\end{table}

\clearpage
\section{DES Latency Distribution: Mean, Median, and p99}
\label{sec:des_appendix}

Full percentile breakdown (mean, median, p99) for the same DeepSeek R1, B300 FP4 disaggregated serving scenario as Figure~\ref{fig:des_p99_b300_fp4}: six 8-GPU prefill workers, one 24-GPU decode worker, Poisson($\lambda{=}250$~req/s), ISL~8192, OSL~1024. Rows correspond to percentile level; columns to TTFT, TPOT, and end-to-end latency. Each datapoint is annotated with its in-flight concurrency (2048, 4096, 6144). Blue: electrical baseline; coral: optical ($4\times$ per-GPU bandwidth).

\begin{figure}[H]
\centering
\includegraphics[width=\textwidth]{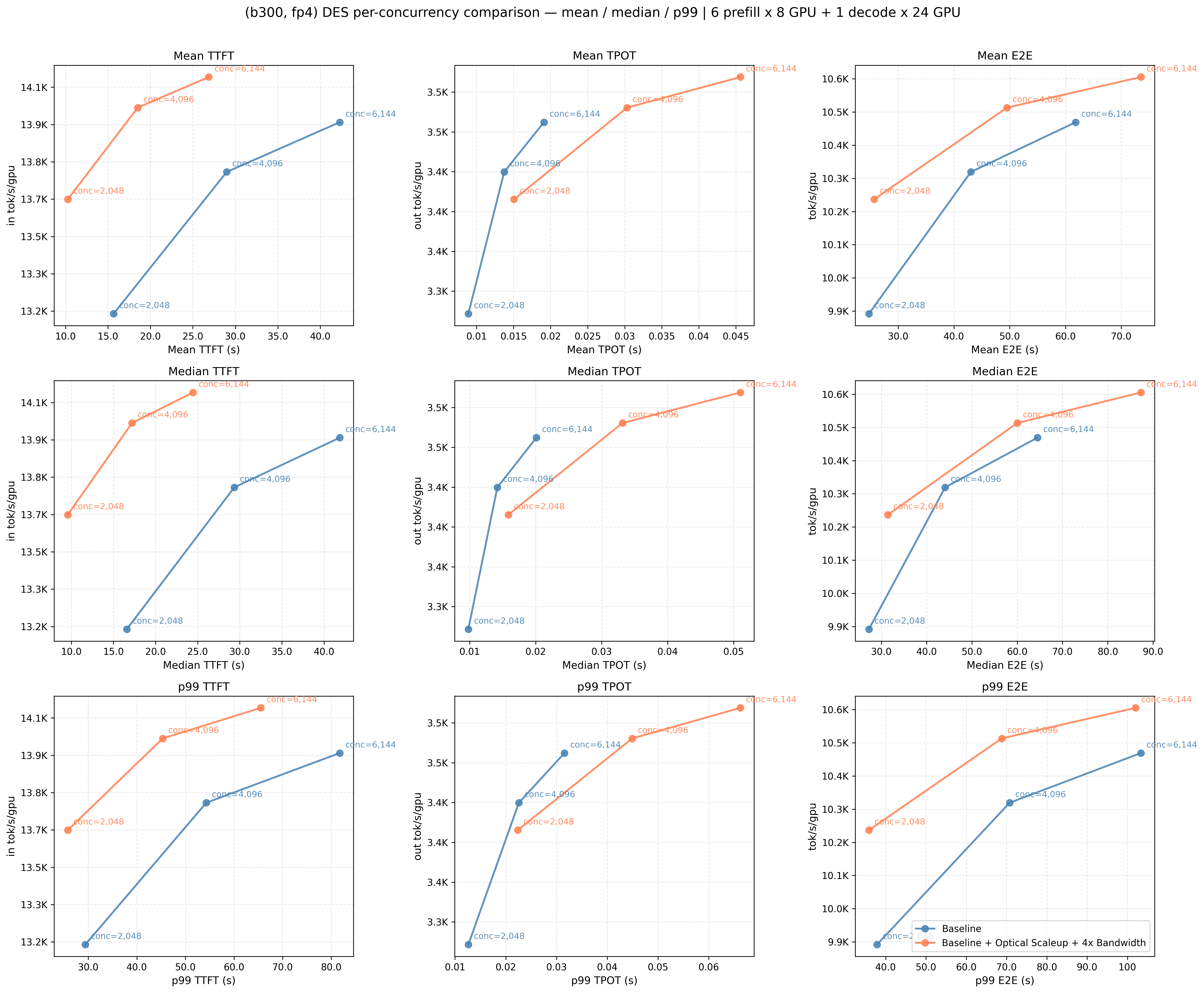}
\caption{DES latency distribution --- DeepSeek R1, B300 FP4, mean / median / p99.}
\label{fig:des_all_b300_fp4}
\end{figure}

\end{document}